\documentclass[prd,preprintnumbers,nofootinbib]{revtex4} 

\usepackage{graphicx}
\usepackage{dcolumn}
\usepackage{bm}
\usepackage{latexsym}
\usepackage{amsfonts}
\usepackage{amssymb}
\usepackage{amsmath}

\newcommand{\eq}{\!\! =\!\!}

\begin{document}

\preprint{IPMU10-0059}

\title{
Trispectrum from Ghost Inflation
}

\author{Keisuke Izumi}
\email{keisuke.izumi@ipmu.jp}
\author{Shinji Mukohyama}
\email{shinji.mukohyama@ipmu.jp}
\affiliation{
 IPMU, The University of Tokyo, Kashiwa, Chiba 277-8582, Japan}

\date{\today}

\begin{abstract}
 Ghost inflation predicts almost scale-invariant primordial cosmological
 perturbations with relatively large non-Gaussianity. The bispectrum is
 known to have a large contribution at the wavenumbers forming an
 equilateral triangle and the corresponding nonlinear parameter
 $f_{NL}^{equil}$ is typically of order $O(10^2)$. In this paper we
 calculate trispectrum from ghost inflation and show that the
 corresponding nonlinear parameter $\tau_{NL}$ is typically of order
 $O(10^4)$. We investigate the shape dependence of the trispectrum and
 see that it has some features different from DBI inflation. Therefore, our
 result may be useful as a template to distinguish ghost inflation from
 other models of inflation by future experiments. 
\end{abstract}

\maketitle

\section{Introduction}
\label{sec:intro}

Almost scale-invariant primordial cosmological perturbations predicted
by inflation fits observational data very well~\cite{Komatsu:2010fb}. 
While this is certainly
a great success of the general idea of inflation, there still remain many
unanswered important questions about inflation.

One of those important questions is how to distinguish different models
of inflation observationally. There are many models of inflation which
are consistent with observational data. It is often thought that tensor
mode fluctuations~\cite{GW} and 
non-Gaussianity~\cite{Maldacena:2002vr,squeeze,equilateral,
ArkaniHamed:2003uz,Holman:2007na,trispectrum,Chen:2009bc,Huang:2006eha,Arroja:2009pd,
Mizuno:2009mv} 
will be useful to distinguish some
of them. Non-Gaussianity is the main subject of the present paper.

While single-field, simple slow-roll inflation predicts negligibly
small non-Gaussianity~\cite{Maldacena:2002vr}, 
it is well known that there are many ways to
generate non-Gaussianities large enough to be detected by near future
experiments. They can be categorized into three by epochs in which
non-Gaussianities are generated: (i) super-horizon, (ii)
horizon-crossing and (iii) sub-horizon epochs. Each of these three types 
has a bispectrum with characteristic dependence on shapes of triangle formed
by wave-vectors. The bispectrum for each type has a large contribution
at (i) squeezed-~\cite{squeeze}, (ii) equilateral-~\cite{Chen:2006nt,equilateral,
ArkaniHamed:2003uz} and (iii) folded-triangle~\cite{Holman:2007na},
respectively. Among them, our interest in the present paper is on the
type (ii), in which large non-Gaussianity is typically due to higher
derivative terms whose importance is enhanced by smallness of the sound
speed. Concrete examples of this type includes k-inflation~\cite{Chen:2006nt}, 
DBI inflation~\cite{Chen:2006nt,equilateral} and ghost 
inflation~\cite{ArkaniHamed:2003uz,ArkaniHamed:2003uy,ArkaniHamed:2005gu,Senatore:2004rj}.

While bispectrum is the leading deviation from Gaussian statistics and
thus a useful tool to distinguish some models of inflation from others, 
trispectrum can also provide additional information about
inflation~\cite{trispectrum,Chen:2009bc,Huang:2006eha,Arroja:2009pd,Mizuno:2009mv,RenauxPetel:2009sj}.
In this paper we shall calculate trispectrum from ghost inflation,
hoping to find ways to distinguish ghost inflation from other
inflationary models which predict similar bispectra. We shall find that
the shape-dependence of trispectrum in ghost inflation shows some
difference from that in DBI
inflation~\cite{Chen:2009bc,Huang:2006eha,Arroja:2009pd,Mizuno:2009mv,RenauxPetel:2009sj}. Therefore, 
the result of the present paper may be useful as a template to 
distinguish different models of inflation by future experiments.

The rest of this paper is organized as follows. In Sec.~\ref{review} 
we  review the ghost inflation and show the powerspectrum and 
the bispectrum in this model. 
In Sec.~\ref{trispectrum} we calculate the trispectrum 
in the ghost inflation. 
In Sec.~\ref{Sum} is devoted to a summary of this paper and discussion. 
In Appendix~\ref{in-in} we give a brief review of the in-in formalism. 
In Appendix~\ref{interactionhamiltonian} we present the method to
construct the interaction Hamiltonian. 
In Appendix~\ref{calculations} we show the details of the calculations.

\section{Review of ghost inflation}
\label{review}

In this section we briefly review the ghost
inflation~\cite{ArkaniHamed:2003uz}, which is an inflationary model in
the ghost condensation~\cite{ArkaniHamed:2003uy,ArkaniHamed:2005gu}. 

\subsection{The Model}
\label{sec:model}

The ghost condensation is the simplest Higgs phase for gravity and
modifies gravity in the infrared. This can be realized if the derivative
of a scalar field obtains a constant, timelike vacuum expectation value
in a maximally symmetric spacetime, either Minkowski or de Sitter 
spacetime. By choosing the time coordinate properly, the vacuum
expectation value of the scalar field can be written in the form
\begin{eqnarray}
\left\langle \phi \right\rangle = M^2 t, 
\label{eqn:dphi-vev}
\end{eqnarray}
where $M$ is a constant. The background energy momentum tensor of the
scalar field is identical to that of a cosmological constant.

Ghost inflation is a model of inflation driven by the scalar field
$\phi$ responsible for ghost condensation. This situation can be
realized, for instance, a la hybrid inflation. We 
introduce another field $\chi$, which is assumed to be massive for
$\phi$ smaller than some critical value $\phi_c$. At the critical point
$\phi=\phi_c$, $\chi$ becomes massless and then becomes tachyonic for
$\phi>\phi_c$ so that inflation ends as in hybrid inflation. An
important point of this model is that the end of inflation depends only
on the value of $\phi$.

In order to make predictions for the CMB anisotropy and compare with
observational data, we need to consider perturbations around the
background. Metric perturbations $h_{\mu\nu}$ are defined as 
\begin{eqnarray}
ds^2 = (-1+h_{tt})dt^2 + 2h_{ti}dtdx^i 
 + (a^2 \delta_{ij}+h_{ij})dx^idx^j,
\end{eqnarray} 
where $a=e^{Ht}$. The scalar field is also decomposed into the
background and perturbation as
\begin{eqnarray}
\phi= M^2 t + \pi. 
\end{eqnarray}
The perturbation $\pi$ is the Nambu-Goldstone boson associated with
ghost condensate, i.e. spontaneous breaking of the time
reparametrization symmetry, and thus we call it 
{\it ghostone}~\cite{ArkaniHamed:2005gu}. Under the general 
infinitesimal diffeomorphism, $x^\mu \to x^\mu + \xi^{\mu}$, 
the perturbations $\pi$ and $h_{\mu\nu}$ are transformed as 
\begin{eqnarray}
&&\pi \mapsto  \pi + M^2 \xi_t, \qquad
h_{tt} \mapsto  h_{tt} -2\partial_t \xi_t, \qquad
h_{ti} \mapsto  h_{ti} -\partial_i \xi_t - a^2 \partial_t (a^{-2}\xi_i) ,
\nonumber \\
&&h_{ij} \mapsto  h_{ij} +2H\xi_t a^2\delta_{ij} -\partial_i \xi_j -\partial_j \xi_i,
\end{eqnarray}
where the indices are raised and lowered by using the background metric.

Let us now construct the quadratic action for $\pi$ in de Sitter
background. We can do this in a systematic
way~\cite{ArkaniHamed:2003uy}. We begin with the unitary gauge,
i.e. $\pi=0$ so that $\phi=M^2t$. This gauge condition still leaves
residual gauge freedom corresponding to the time-dependent spatial
diffeomorphism: 
\begin{eqnarray}
 t \mapsto t, \qquad x^i \mapsto x^i + \xi^i(t,x). 
\end{eqnarray}
Under this transformation, the gravitational perturbations $h_{\mu\nu}$ 
are transformed as 
\begin{eqnarray}
h_{tt} \mapsto  h_{tt}, \qquad
h_{ti} \mapsto  h_{ti} - a^2 \partial_t (a^{-2}\xi_i) , \qquad
h_{ij} \mapsto  h_{ij}  -\partial_i \xi_j -\partial_j \xi_i.
\end{eqnarray}
Since the quadratic action must be invariant under this transformation, 
it should be constructed from the following three terms:
$\int dtdx^3a^3h_{tt}^2$, $\int dtdx^3a^3K^{ij}K_{ij}$
and $\int dtdx^3a^3K^2$, where 
\begin{eqnarray}
 K_{ij} \equiv \frac{1}{2}
 \left[
  a^2\partial_t (a^{-2}h_{ij}) 
  -\partial_i h_{tj}-\partial_j h_{ti}
 \right],
\end{eqnarray}
$K^{ij}=a^{-4}\delta^{ik}\delta^{jl}K_{kl}$ and
$K=a^{-2}\delta^{ij}K_{ij}$.

We can then obtain the quadratic action in general gauge by undoing the
unitary gauge. This is achieved by performing the spontaneously broken
diffeomorphism $\xi_t=M^{-2}\pi$, under which 
\begin{eqnarray}
&&h_{tt} \to h_{tt}-2M^{-2}\partial_t\pi, \\
&&K_{ij} \to K_{ij} + M^{-2}
 (Ha^2\delta_{ij}\partial_t\pi +\partial_i\partial_j\pi). 
\end{eqnarray}
After taking the decoupling limit $M/M_{Pl}\to 0$ and thus dropping
$h_{\mu\nu}$ in the quadratic action, we obtain the leading quadratic
action for $\pi$, 
\begin{eqnarray}
S_2=\int dt dx^3 a^3 \left[ \frac{1}{2} (\partial_t \pi)^2 
- \frac{\alpha_1}{2 M^2} \left(\frac{\vec{\nabla}^2}{a^2}
			\pi+3H\partial_t\pi\right)^2
- \frac{\alpha_2}{2 M^2} 
\left(\frac{\vec{\nabla}^i\vec{\nabla}^j}{a^2}
 \pi+H\delta^{ij}\partial_t\pi\right) 
\left(\frac{\vec{\nabla}_i\vec{\nabla}_j}{a^2}
 \pi+H\delta_{ij}\partial_t\pi\right) 
\right] , 
\label{eq:2piaction-pre}
\end{eqnarray}
where $\alpha_1$ and $\alpha_2$ are constants of order unity and we have normalized
$\pi$. Here, $\vec{\nabla}^i=\delta^{ij}\vec{\nabla}_j$ and
$\vec{\nabla}^2=\vec{\nabla}^i\vec{\nabla}_i$. 
For $H^2/M^2\ll 1$, as we shall justify soon, we can drop the terms
depending on $H\partial_t\pi$ and thus the quadratic action is reduced
to 
\begin{eqnarray}
S_2=\int dt dx^3 a^3 \left[ \frac{1}{2} (\partial_t \pi)^2 
- \frac{\alpha}{2 M^2}
\left(\frac{\vec{\nabla}^2}{a^2}\pi\right)^2 \right], 
\label{eq:2piaction}
\end{eqnarray}
where $\alpha=\alpha_1+\alpha_2$. 
Therefore, there is no $(\vec{\nabla} \pi)^2$ term in the quadratic
action for $\pi$ and the lowest spatial derivative term is 
$a^{-4}(\vec{\nabla}^2 \pi)^2$. In particular, the dispersion relation
for the ghostone $\pi$ is 
\begin{eqnarray}
 \omega^2 = \frac{\alpha}{M^2}\frac{k^4}{a^4}. 
\end{eqnarray}
This implies that the fluctuation of $\pi$ oscillates in the regime
$k^2/M\gg Ha$ while it freezes for $k^2/M\ll Ha$. Moreover, because of
the shift symmetry, interaction terms in the $\pi$ action always include
derivatives and thus are suppressed outside the sound horizon, i.e. for
$k^2/M\ll Ha$. Therefore, it is interactions during the epoch around the
sound horizon crossing $k^2/M\sim Ha$ that essentially determines the
bispectrum and higher-order correlation functions of cosmological
perturbations. Near the sound horizon crossing, we have 
$\omega\sim (k/a)^2/M\sim H$ and thus 
$(k/a)^2\sim MH\gg H^2\sim H\omega$ if $H/M\ll 1$. We shall see below
that the COBE normalization for curvature perturbations fixes $H/M$ to a
small value (\ref{eqn:HoverM}). This is the reason why in
(\ref{eq:2piaction-pre}), $H\partial_t\pi$ can be neglected compared
with $a^{-2}\vec{\nabla}\vec{\nabla}\pi$.

In order to see the leading cubic and quartic interactions, we should
identify the scaling dimensions of time coordinate $t$, spatial coordinates 
$x^i$ and the ghostone $\pi$. Suppose that energy $E$ is scaled by a
factor of $s$ as $E \to s E$ and that time $t$ is scaled as 
$t \to s^{-1} t$. Since we know from the quadratic effective action for
$\pi$ (\ref{eq:2piaction}) that the dispersion relation is 
$\omega^2 \propto k^4$,  we have to scale $k$ as $k\to s^{1/2} k$ or $x$
as $x\to s^{-1/2} x$. Then, by demanding that the action is invariant
under scaling, the scaling dimension of $\pi$ is determined to be $1/4$:  
$\pi\to s^{1/4} \pi$. With the use of these scaling dimensions, we can
identify the leading cubic and quartic operators.

When we introduced (\ref{eqn:dphi-vev}) as a consistent background, we
implicitly assumed the shift symmetry, i.e. the invariance of the theory
under a constant shift of the scalar field
$\phi\to\phi+\mbox{const}$. Because of the shift symmetry, the
fluctuation $\pi$ appears always with derivatives.

We further assume that the theory is invariant under the $Z_2$
transformation $\phi \to -\phi$. This corresponds to the simultaneous
change of the signs of $t$ and $\pi$: 
\begin{eqnarray}
t \to -t \qquad \mbox{and} \qquad \pi \to -\pi. 
\label{eq:Z2}
\end{eqnarray} 
For example, $\partial_t \pi\vec{\nabla}^2 \pi$ (without being multiplied by
$H$) is forbidden by this symmetry.

Let us now seek the leading cubic operator. The shift symmetry
tells us that a cubic term has at least three derivatives. Because of
the $Z_2$-symmetry (\ref{eq:Z2}), a cubic term should have one or three
time derivatives. Combining these with the fact that the scaling
dimension of spatial derivative is lower than that of time derivative,
we conclude that the leading cubic operator is
\begin{eqnarray}
S_3= - \frac{\beta}{2M^2} \int dt dx^3 a^3  \partial_t \pi 
\frac{(\vec{\nabla} \pi)^2}{a^2},
\label{S3}
\end{eqnarray}
where $\beta$ is a constant of unity. This operator has the scaling
dimension $1/4$. Similarly, the leading quartic operator is 
\begin{eqnarray}
S_4= -\frac{\gamma}{8M^4} \int dt dx^3 a^3  
\frac{(\vec{\nabla} \pi)^4}{a^4},
\end{eqnarray}
where $\gamma$ is a constant of order unity, and has the scaling
dimension $1/2$.

\subsection{Powerspectrum}

In this subsection we review the power spectrum of the curvature
perturbation generated by ghost inflation.

The scale $M$, which plays the role of ultraviolet cutoff of the
effective field theory, is constrained by various
observations. The strongest constraint is from the twinkling by 
lensing which bounds $M \lesssim
100$GeV~\cite{ArkaniHamed:2005gu}~\footnote{
It is known that the ghostone behaves like dark matter. If we suppose
that the ghostone is responsible for all dark matter then the structure
formation gives a lower bound on $M$: 
$M\gtrsim 10$eV~\cite{Furukawa:2010gr}.}.
Since this scale is much lower than the Planck scale $M_{\rm Pl}$,
decoupling limit is a good approximation. Therefore, in order to
calculate the power spectrum of ghostone $\pi$ we can simply study
the $\pi$ action without coupling to gravity. We shall later relate
$\pi$ to the curvature perturbation $\zeta$.

The equation of motion for $\pi$ in the linearized level is 
\begin{eqnarray}
{u_{\bf k}}'' - \frac{2}{\eta^2}u_{\bf k} + 
\frac{\alpha k^4H^2 \eta^2}{M^2}u_{\bf k}
=0,
\end{eqnarray}
where $u_{\bf k}= a\pi_{\bf k}$, ${\bf k}$ is a comoving wavevector and
$k=|{\bf k}|$. Here, we have introduced the conformal time
\begin{equation}
 \eta = -\frac{H}{a}
\end{equation}
so that $dt=ad\eta$, and a prime represents derivative with respect to
$\eta$.

We quantize $u$ as usual 
\begin{eqnarray}
u_{\bf k}(t) = w_k(t)a_{\bf k}+w_k^*(t)a_{-{\bf k}}^\dagger,
\end{eqnarray}
where $a_{\bf k}$ and $a_{-{\bf k}}^\dagger$ are
annihilation and creation operators satisfying 
$[a_{\bf k},a_{{\bf k}'}^\dagger]=(2\pi)^3\delta^3({\bf k}-{\bf k}')$. 
Choosing $w_k(t)$ so that it corresponds to the correct mode function in
flat spacetime for very short wavelength, we have 
\begin{eqnarray}
w_{k}(\eta)=\sqrt{\frac{\pi}{8}}(-\eta)^{1/2}
H_{3/4}^{(1)}(q \eta^2), 
\end{eqnarray}
where
\begin{eqnarray}
q \equiv \frac{\sqrt{\alpha} H k^2}{2M}.
\end{eqnarray}

The observed fluctuations in the CMB are generated quantum mecanically
and streched by the exponential expansion during the inflationary
stage of the universe. Their wavelengths exceed the Hubble horizon and,
at the end of the inflation, are much longer than the Hubble horizon
scale. This means 
\begin{eqnarray}
 k|\eta_e| \ll 1,
\end{eqnarray}
where $k$ is the wavenumber of the mode of interest and $\eta_e$ is the
conformal time at the end of inflation. Therefore, it is a good
approximation to take the limit $\eta_e\to 0$.

The power spectrum of the ghostone $\pi$ is obtained as 
\begin{eqnarray}
{\cal P}_\pi= \left.\frac{k^3}{2\pi^2}\left|\frac{w_k}{a}\right|^2
\right|_{\eta \to 0} 
=  \frac{H^{1/2}M^{3/2}}{\pi (\Gamma(1/4))^2 \alpha^{3/4}}.
\end{eqnarray}
The gauge invariant curvature perturbation $\zeta$
~\cite{Bardeen:1983qw} is related to $\pi$ as 
\begin{eqnarray}
\zeta = -\frac{H}{\dot \phi} \pi = -\frac{H}{M^2}\pi. 
\label{zeta-pi-relation}
\end{eqnarray}
Therefore, we obtain the power spectrum of the primordial curvature
perturbation as~~\cite{ArkaniHamed:2003uz}  
\begin{eqnarray}
{\cal P}_\zeta=\left(-\frac{H}{M^2}\right)^2 {\cal P}_\pi
= \frac{1}{\pi (\Gamma(1/4))^2 \alpha^{3/4}}
\left( \frac{H}{M} \right)^{5/2}.
\label{H/M}
\end{eqnarray}
The COBE normalization sets 
${\cal P}_\zeta^{1/2} \simeq 4.8 \times10^{-5}$, and thus
\begin{eqnarray}
 \frac{H}{M} \simeq (1.6\times 10^{-3})\times\alpha^{3/10} \ll 1.
  \label{eqn:HoverM}
\end{eqnarray}

\subsection{Bispectrum}
\label{subsec:bispectrum}

Let us now review the bispectrum of the curvature perturbation generated by
ghost inflation.

As in the previous subsection, we calculate a correlation function for
$\pi$ in the decoupling limit and then relate the ghostone $\pi$ to the 
curvature perturbation $\zeta$. For the calculation of the bispectrum,
nonlinearity in the relation between $\pi$ and $\zeta$, if any, should
be taken into account. Fortunately, in our model of ghost inflation the
simple linear relation (\ref{zeta-pi-relation}) is exact and there is no
nonlinearity involved in the relation between $\pi$ and $\zeta$,
essentially because the background $\phi$ is linear in $t$. Therefore,
there are no additional nonlinear effects due to transformation of $\pi$
to $\zeta$, and the bispectrum of $\zeta$ is simply proportional to that
of $\pi$.

As we have already seen in subsection~\ref{sec:model}, the leading cubic
interaction in the $\pi$ action is 
\begin{eqnarray}
-\frac{\beta}{2M^2} \int dt dx^3 a^3  \partial_t \pi 
 \frac{(\vec{\nabla} \pi)^2}{a^2}.
\end{eqnarray}
The corresponding operator in the Hamiltonian density is 
\begin{eqnarray}
{\cal H}_{int,3}= a^3  \frac{\beta}{2M^2} \partial_t \pi 
\frac{(\vec{\nabla} \pi)^2}{a^2}.
\end{eqnarray}
Using the in-in formalism, we can calculate the tree-level bispectrum 
of ghostone $\pi$, $\langle \pi \pi \pi \rangle$, as 
\begin{eqnarray}
\langle \pi_{{\bf k}_1}(t) \pi_{{\bf k}_2}(t) 
\pi_{{\bf k}_3}(t) \rangle \Bigr|_{t\to \infty}&\eq& 
-i \int_{-\infty}^{t} dt' 
\left\langle\left[
\pi_{{\bf k}_1}(t) \pi_{{\bf k}_2}(t) 
\pi_{{\bf k}_3}(t), \int dx^3 {\cal H}_{int,3}(t')
\right]
\right\rangle \Bigr|_{t\to \infty}
\nonumber\\
&\eq&
-\frac{4\sqrt{2}\pi^{3/2} H M^2 \beta}
{\bigl(\Gamma(1/4)\bigr)^3 \alpha^2} (2 \pi^3) \delta^3
\left( \sum_i {\bf k}_i \right) \prod _i k_i^{-3}
\nonumber\\
&&\qquad
\Re\left[
\int _{-\infty}^0 d\eta \, \eta^{-1} 
F(\eta) F\left( \frac{k_2}{k_1} \eta \right)
F\left( \frac{k_3}{k_1} \eta \right) k_3 ({\bf k}_1\cdot{\bf k}_2)
+ \mbox{symm.}
\right] ,
\end{eqnarray}
where 
\begin{eqnarray}
F(\eta)=\sqrt{\frac{\pi}{8}}(\eta)^{3/2}H^{(1)}_{3/4}(\eta^2/2).
\end{eqnarray}
Translating this into the bispectrum of $\zeta$, we obtain 
\begin{eqnarray}
\langle \zeta_{{\bf k}_1}\zeta_{{\bf k}_2}\zeta_{{\bf k}_3} \rangle 
&\eq&
-\frac{4\sqrt{2}\pi^{3/2}  \beta}
{\bigl(\Gamma(1/4)\bigr)^3 \alpha^2}
\left(\frac{H}{M}\right)^4
 (2 \pi^3) \delta^3
\left( \sum_i {\bf k}_i \right) \prod _i k_i^{-3}
\nonumber\\
&&\qquad
\Re\left[
\int _{-\infty}^0 d\eta \, \eta^{-1} 
F(\eta) F\left( \frac{k_2}{k_1} \right)
F\left( \frac{k_3}{k_1} \right) k_3 ({\bf k}_1\cdot{\bf k}_2)
+ \mbox{symm.}
\right] .
\label{eqn:bispectrum-zeta}
\end{eqnarray}

It is known that the bispectrum from ghost inflation has a large
contribution at equilateral triangles formed by three vectors ${\bf
k}_1$, ${\bf k}_2$ and 
${\bf k}_3=-{\bf k}_1-{\bf k}_2$~\cite{Senatore:2004rj}. 
The nonlinear parameter for equilateral configuration 
$f^{equil}_{NL}$ is defined 
as~\cite{Creminelli:2005hu,Komatsu:2010fb}
\begin{eqnarray}
\left.\left\langle \zeta_{{\bf k}_1}\zeta_{{\bf k}_2}
\zeta_{{\bf k}_3}\right\rangle\right|_{k_1=k_2=k_3=k}=
(2\pi)^3 \delta^3(\sum_i {\bf k}_i) 
\cdot
\frac{6}{5}f^{equil}_{NL}
\cdot
3\left(\frac{2\pi^2{\cal P}_\zeta}{k^3}\right)^2.
\end{eqnarray}
The value corresponding to the result (\ref{eqn:bispectrum-zeta}) 
is~\cite{ArkaniHamed:2003uz} 
\begin{eqnarray}
f_{NL}^{equil}\simeq  85 \cdot \beta \cdot \alpha^{-4/5}.
\end{eqnarray}
The WMAP 7-year data bounds $f_{NL}^{equil}$~\cite{Komatsu:2010fb} as 
\begin{eqnarray}
f_{NL}^{equil} = 26 \pm 140 (68\%\ \mbox{CL}).
\end{eqnarray}
It is natural to take $\alpha\sim \beta\sim 1$ as they are
dimensionless parameters. In this case, the ghost inflation is not
excluded by the WMAP 7-year constraint but can be detected in the near 
future by, say, the PLANCK satellite.

\section{Trispectrum}
\label{trispectrum}

In this section we show the four-point function of late-time
curvature perturbation $\zeta$ from ghost inflation. Details of
calculations are presented in Appendix~\ref{calculations}.

As we have already seen in subsection~\ref{sec:model}, the leading
quartic interaction in the $\pi$ action is 
\begin{eqnarray}
 -\frac{\gamma}{8M^4} \int dt dx^3 a^3  
\frac{(\vec{\nabla} \pi)^4}{a^4},
\end{eqnarray}
As shown in Appendix~\ref{interactionhamiltonian}, the corresponding
operator in the Hamiltonian density is 
\begin{eqnarray}
{\cal H}_{int,4}=   \frac{\tilde \gamma}{8M^4} a^{-1}(\vec{\nabla} \pi)^4 ,
 \quad \tilde \gamma = \gamma +2 \beta^2,
\end{eqnarray}
where $\beta$ is the dimensionless coefficient of the leading cubic 
interaction.

\begin{figure}[t]
\begin{minipage}{0.49\textwidth}
  \begin{center}
    \includegraphics[keepaspectratio=true,height=40mm]{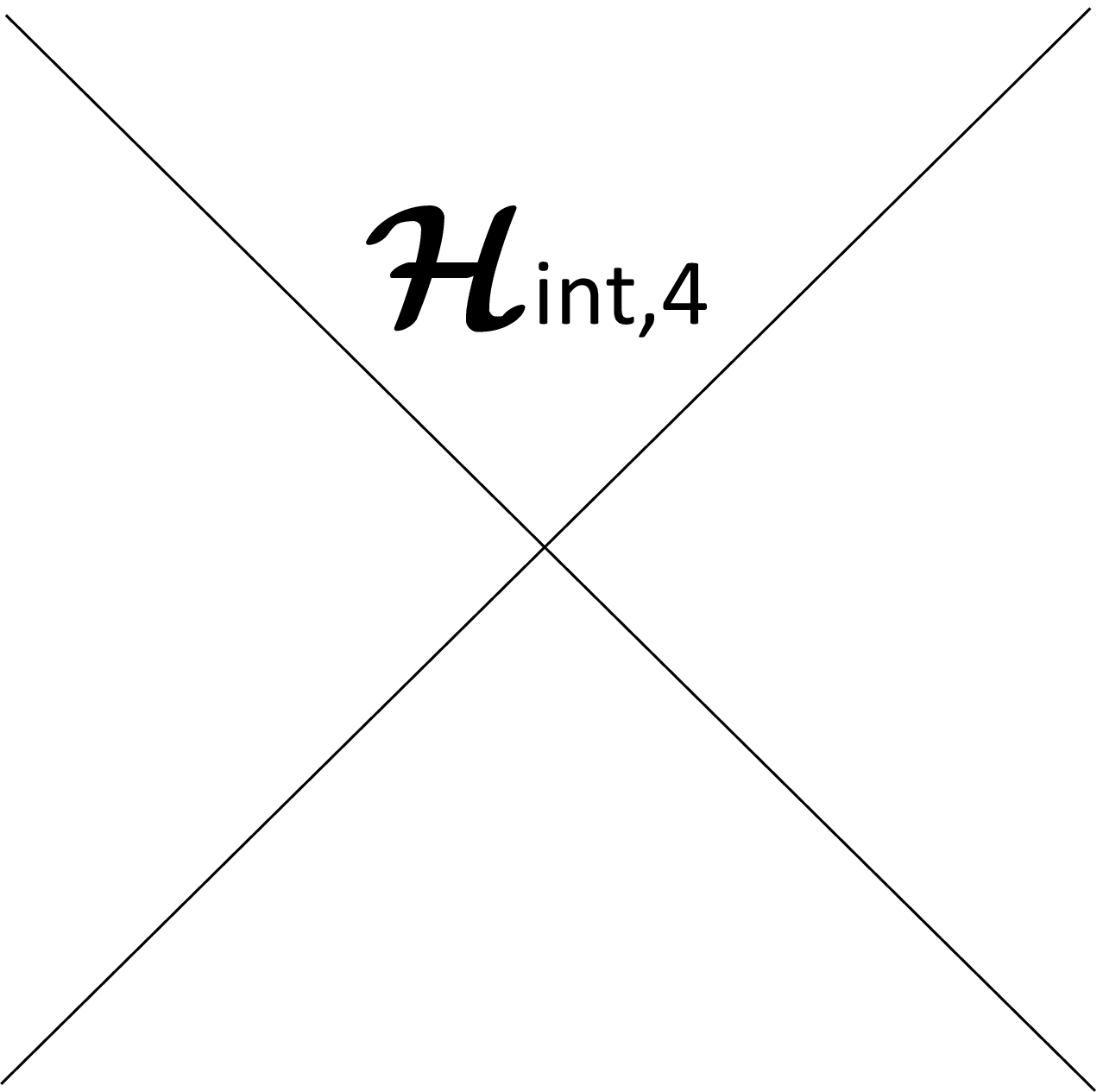}
  \end{center}
  \caption{The diagram for the contact term contribution.}  
  \label{fig:4-point.eps}
  \end{minipage}
 \begin{minipage}{0.49\textwidth}
  \begin{center}
    \includegraphics[keepaspectratio=true,height=40mm]{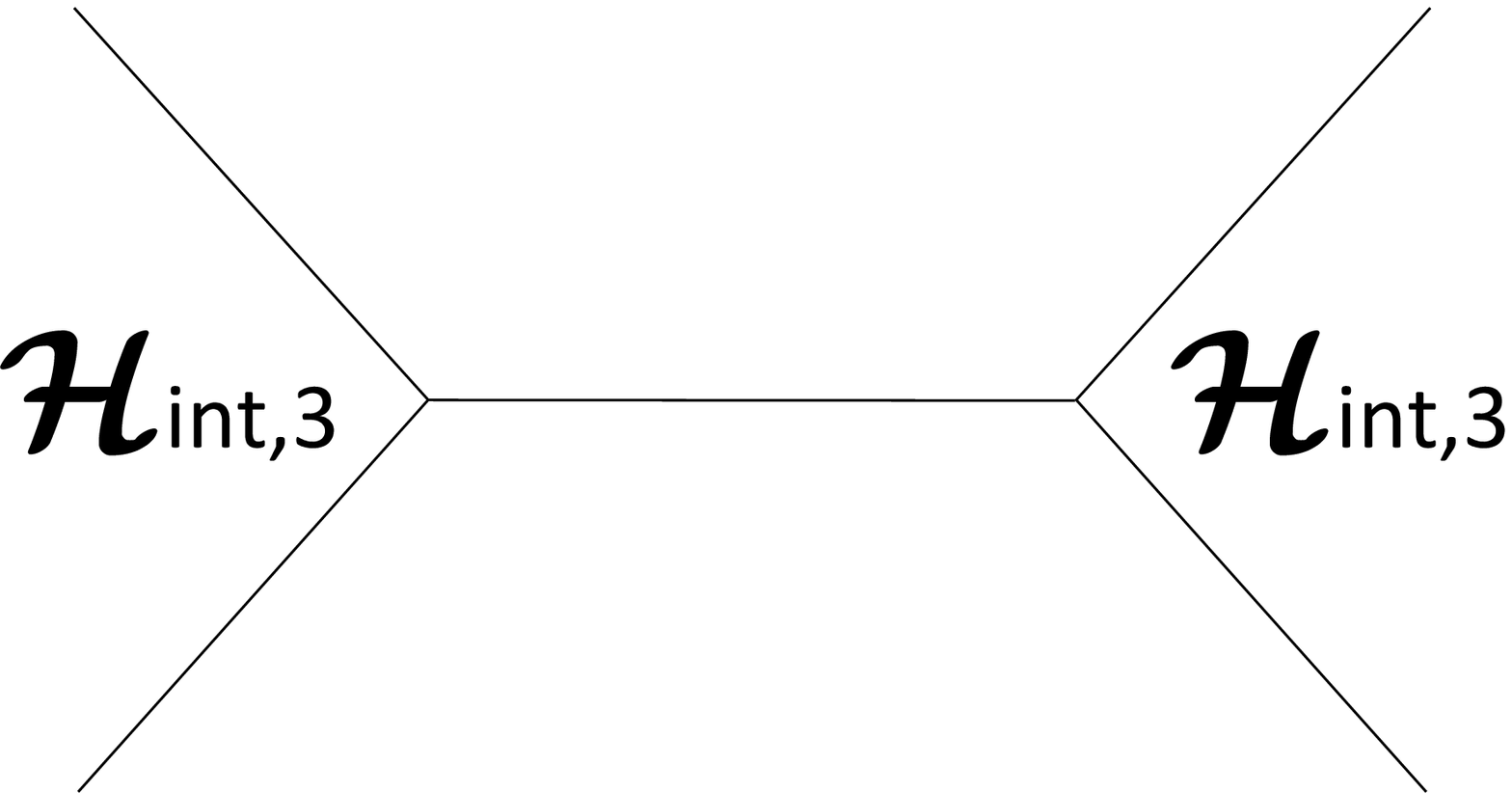}
  \end{center}
  \caption{The diagram for the scalar exchange contribution.}
    \label{fig:3-point.eps}
     \end{minipage}
\end{figure}

There are two contributions to the four-point function: a contact term
contribution and a scalar exchange contribution. According to the in-in
formalism (See Appendix~\ref{in-in}), the leading contributions to the
four-point function are written as 
\begin{eqnarray}
&&\left\langle 
\pi_{\bf k_1}(t)\pi_{\bf k_2}(t)\pi_{\bf k_3}(t)\pi_{\bf k_4}(t)
\right\rangle \nonumber\\
&&\qquad\equiv
i \int_{-\infty}^t dt' \left\langle
\left[ \int d^3x' {\cal H}_{int,4}(t',x'),
\pi_{k_1}(t)\pi_{k_2}(t)\pi_{k_3}(t)\pi_{k_4}(t)\right]
\right\rangle 
\nonumber\\
&&\qquad\qquad\qquad
- \int^t_{-\infty} dt_2 \int^{t_2}_{-\infty} dt_1
\bigg\langle \biggl[ \int dx_1^3 {\cal H}_{int,3}(t_1,x_1), 
\Bigl[\int dx_2^3 {\cal H}_{int,3}(t_2,x_2), 
\pi_{\bf k_1}\pi_{\bf k_2}\pi_{\bf k_3}\pi_{\bf k_4} \Bigr]\biggr] \bigg\rangle.
\label{eq:4-point}
\end{eqnarray}
The first term in the right hand side of eq.(\ref{eq:4-point}) is called
the contact term contribution and discribed in
Fig.~\ref{fig:4-point.eps} as a four-point vertex. 
The second term is called the scalar exchange contribution whose diagram
is shown in Fig.~\ref{fig:3-point.eps}. This includes two three-point
vertices. We calculate these contributions of the four-point function
separately.

\subsection{contact term contribution}

Let us consider the contact term contribution to the four-point function. 
The contact term contribution to the four-point function is defined as 
\begin{eqnarray}
\left\langle 
\pi_{\bf k_1}(t)\pi_{\bf k_2}(t)\pi_{\bf k_3}(t)\pi_{\bf k_4}(t)
\right\rangle _{cc}
\equiv
i \int_{-\infty}^t dt' \left\langle
\left[\frac{\tilde \gamma}{8M^4} \int d^3x' \frac{ (\vec{\nabla} \pi(t',x'))^4}{a(t')},
\pi_{k_1}(t)\pi_{k_2}(t)\pi_{k_3}(t)\pi_{k_4}(t)\right]
\right\rangle ,
\end{eqnarray}
which depends on $\tilde \gamma$. This parameter $\tilde \gamma$ 
does not affect the leading-order bispectrum and thus the four-point
function provides independent information about the effective action.

We define contact term contributions to trispectrums 
$T_{\pi ,cc}(\eta,{\bf k_1},{\bf k_2},{\bf k_3},{\bf k_4})$ 
and $T_{\zeta ,cc}(\eta,{\bf k_1},{\bf k_2},{\bf k_3},{\bf k_4})$ 
as
\begin{eqnarray}
&&\left\langle 
\pi_{\bf k_1}(t)\pi_{\bf k_2}(t)\pi_{\bf k_3}(t)\pi_{\bf k_4}(t)
\right\rangle _{cc}  \equiv (2\pi)^3 \delta^3
({\bf k_1}+{\bf k_2}+{\bf k_3}+{\bf k_4}) 
T_{\pi ,cc}(\eta,{\bf k_1},{\bf k_2},{\bf k_3},{\bf k_4}).
\\
&&\left\langle 
\zeta_{\bf k_1}(t)\zeta_{\bf k_2}(t)\zeta_{\bf k_3}(t)\zeta_{\bf k_4}(t)
\right\rangle _{cc}  \equiv (2\pi)^3 \delta^3
({\bf k_1}+{\bf k_2}+{\bf k_3}+{\bf k_4}) 
T_{\zeta ,cc}(\eta,{\bf k_1},{\bf k_2},{\bf k_3},{\bf k_4}),
\end{eqnarray}
where $\eta$ is comoving time. As mentioned at the beginning of
subsection~\ref{subsec:bispectrum}, the simple linear relation
(\ref{zeta-pi-relation}) holds and there is no nonlinearity involved 
in the relation between $\pi$ and $\zeta$. Therefore, we have the
following simple relation between two trispectrums. 
\begin{eqnarray}
T_{\zeta ,cc}(\eta,{\bf k_1},{\bf k_2},{\bf k_3},{\bf k_4})
=\left(\frac{H}{M^2} \right)^4 
T_{\pi ,cc}(\eta,{\bf k_1},{\bf k_2},{\bf k_3},{\bf k_4}).
\end{eqnarray}

As shown in Appendix~\ref{calculations} and presented as (\ref{eq:triccpi}),
the contact term contribution to the trispectrum of curvaure
perturbation is 
\begin{eqnarray}
&&T_{\zeta ,cc}(\eta=0,{\bf k_1},{\bf k_2},{\bf k_3},{\bf k_4})
\nonumber\\
&&\qquad= 
\frac{\tilde \gamma }{2^9 }\left(\frac{H}{M}\right)^{12} 
\left( \frac{\pi}{\Gamma(1/4)}\right)^4
\left( q_1 q_2 q_3 q_4 \right)^{-3/4}
\Bigl(
({\bf k_1} \cdot {\bf k_2})({\bf k_3} \cdot {\bf k_4}) +(\mbox{23 terms})
\Bigr)
\nonumber\\
&&\qquad\qquad\qquad\times
\Re\biggl\{ i
\int^0_{-\infty}d\eta' 
\left((-\eta')^{3/2}H_{3/4}^{(1)}(q_1 {\eta'}^2) \right)
\left((-\eta')^{3/2}H_{3/4}^{(1)}(q_2 {\eta'}^2) \right)
\nonumber\\ 
&&\qquad\qquad\qquad\qquad\qquad\qquad\qquad\qquad\qquad
\left((-\eta')^{3/2}H_{3/4}^{(1)}(q_3 {\eta'}^2) \right)
\left((-\eta')^{3/2}H_{3/4}^{(1)}(q_4 {\eta'}^2) \right)
\biggr\} ,
\end{eqnarray}
where 
\begin{eqnarray}
&&q_i \equiv \frac{\sqrt{\alpha} H k_i^2}{2M}.
\label{qi}
\end{eqnarray}

\begin{figure}[t]
\begin{minipage}{0.42\textwidth}
  \begin{center}
    \includegraphics[keepaspectratio=true,height=60mm]{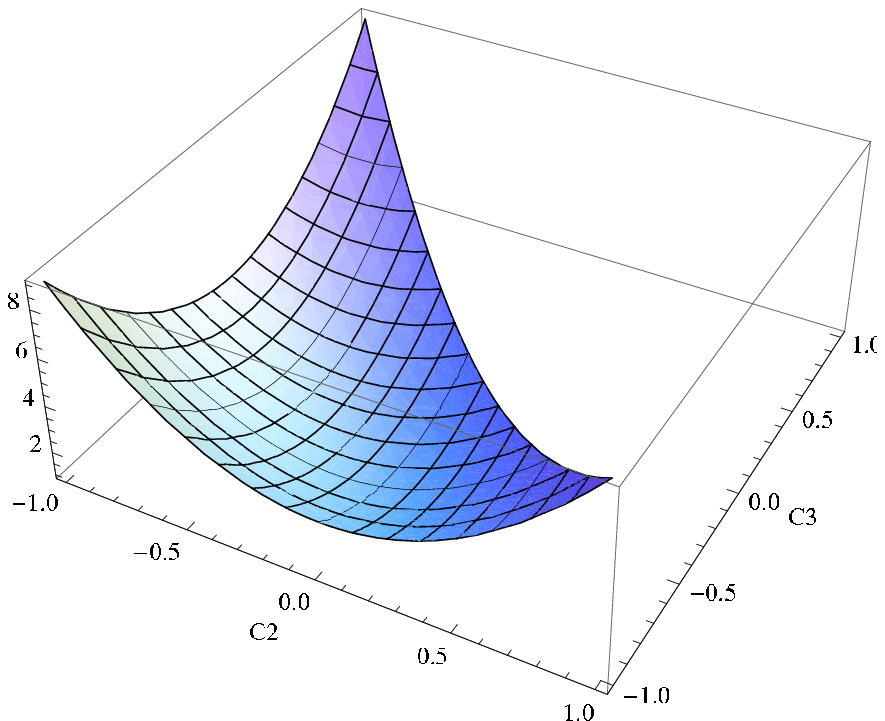}
  \end{center}
  \caption{
 This plot shows the dependence of $T_{\zeta , cc}(k,C_2,C_3,C_4)$ 
 on $C_2$ and $C_3$, where $C_4=-1-C_2-C_3$. 
 The horizontal axis is rescaled by the value at $C_2=C_3=-1/3$.} 
  \label{fig:Tcc.eps}
  \end{minipage} 
  \begin{minipage}{0.05\textwidth} 
  \begin{center}
  \end{center}
  \end{minipage} 
  \begin{minipage}{0.42\textwidth}
  \begin{center}
    \includegraphics[keepaspectratio=true,height=60mm]{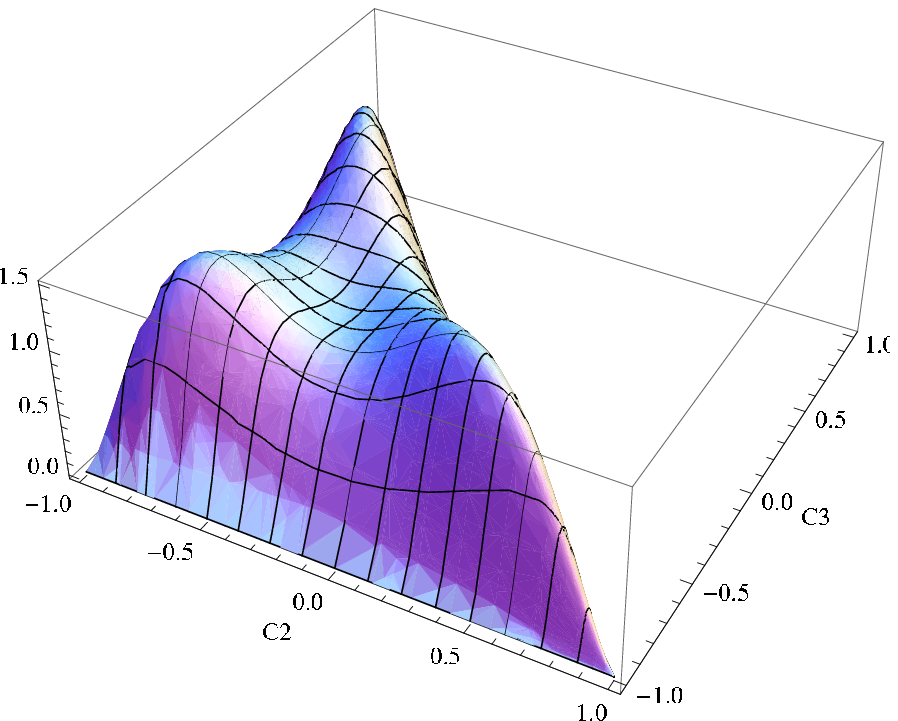}
  \end{center}
  \caption{
  This plot shows the dependence of 
   $\tau_{NL}^{cc}(k,C_2,C_3,C_4)$ on $C_2$ and $C_3$, where
   $C_4=-1-C_2-C_3$. 
   The horizontal axis is rescaled by the value at $C_2=C_3=-1/3$.}
  \label{fig:cc.eps}
\end{minipage}   
\end{figure}

The trispectrum depends on four $3$-momenta ${\bf k}_i$ ($i=1,2,3,4$)
satisfying the constraint $\sum_i{\bf k}_i=0$ representing the momentum 
conservation. Assuming homogeneity and isotropy of the background, the
independent variables are four amplitudes of the momenta, 
$k_i=|{\bf k}_i|$, and two angles between momemta, 
${\bf k_1}\cdot {\bf k_2}/(k_1k_2)$ and 
${\bf k_1}\cdot {\bf k_3}/(k_1k_3)$. Although it is ideal to investigate
the dependence of the trispectrum on all six parameters, it is somehow
complicated. Fortunately, for the purpose of showing some differences
between ghost inflation and other models of inflation such as DBI 
inflation, it is sufficient to investigate the dependence on a subset of
six parameters. Following previous
works~\cite{Arroja:2009pd,Chen:2009bc} on the trispectrum from DBI
inflation, we consider the equilateral case where
\begin{eqnarray}
k_1=k_2=k_3=k_4=k,  \label{eq:k-equal}
\end{eqnarray}
and the remaining independent variables are the two angles described
above. As shown in (\ref{eq:Tzeta}), the equilateral trispectrum of the
curvaure perturbation is 
\begin{eqnarray}
T_{\zeta , cc}(k,C_2,C_3,C_4)= -2.215\times 10^{-17} \times\left(
\frac{\bigl( {\cal P}_\zeta(k)\bigr)^{1/2}}{4.8 \times 10^{-5}}\right)^{22/5}
\frac{\tilde \gamma}{\alpha ^{8/5}} \left( \sum_{i=2,3,4} C_i^2 \right) k^{-9},
\end{eqnarray}
where
\begin{eqnarray}
\frac{{\bf k_1}\cdot {\bf k_2}}{k^2}=\frac{{\bf k_3}\cdot {\bf k_4}}{k^2}
\equiv C_2, \qquad 
\frac{{\bf k_1}\cdot {\bf k_3}}{k^2}=\frac{{\bf k_2}\cdot {\bf k_4}}{k^2}
\equiv C_3, \qquad 
\frac{{\bf k_1}\cdot {\bf k_4}}{k^2}=\frac{{\bf k_2}\cdot {\bf k_3}}{k^2}
\equiv C_4, \qquad 
\label{eq:C_i}
\end{eqnarray}
and $C_i$ ($i=2,3,4$) are constrained by the momentum conservation as 
\begin{eqnarray}
1+\sum_{i=2,3,4} C_i =0.
\label{momcon}
\end{eqnarray}
Fig~\ref{fig:Tcc.eps} shows the dependence of the contact term
contribution to the equilateral trispectrum on the variables $C_2$ and
$C_3$. In the most symmetric case where $C_2=C_3=C_4=-1/3$, the
trispectrum has the smallest absolute value, 
\begin{eqnarray}
T_{\zeta , cc}(k,C_2=-\frac{1}{3},C_3=-\frac{1}{3},C_4=-\frac{1}{3})
= -1.340 \times 10^{-18} \left(
\frac{\bigl( {\cal P}_\zeta(k)\bigr)^{1/2}}{4.8 \times 10^{-5}}\right)^{22/5}
\frac{\tilde \gamma}{\alpha ^{8/5}} k^{-9}.
\end{eqnarray}

Following Ref.~\cite{Arroja:2009pd}, we define the nonlinear parameter
$\tau_{NL}^{cc}$ as 
\begin{eqnarray}
\tau_{NL}^{cc}({\bf k_1},{\bf k_2},{\bf k_3},{\bf k_4})&\!\! \equiv\!\!& 
\frac{T_{\zeta, cc}({\bf k_1},{\bf k_2},{\bf k_3},{\bf k_4})}
{(2 \pi^2 {{\cal P_{\zeta}}( k_0 )})^3}
\prod_{i=1}^4 k_i^3
\nonumber\\
&&\times \left[ 
\left(k_1^3 k_2^3+k_3^3 k_4^3\right)\left(k_{13}^{-3}+k_{14}^{-3} \right) 
+\left(k_1^3 k_4^3+k_2^3 k_3^3\right)\left(k_{12}^{-3}+k_{13}^{-3} \right) 
+\left(k_1^3 k_3^3+k_2^3 k_4^3\right)\left(k_{12}^{-3}+k_{14}^{-3} \right) 
\right]^{-1}, \nonumber \\
&& \qquad
\end{eqnarray}
where $k_0$ is the wavelength coresponding to the present Hubble horizon size
and $k_{ij}=|{\bf k}_i+{\bf k}_j|$. 
Since ${{\cal P_{\zeta}}( k )}$ is almost scale invariant, 
we do not distinguish between ${{\cal P_{\zeta}}( k )}$ and 
${{\cal P_{\zeta}}( k_0 )}$ hereafter.
In the equilateral case (\ref{eq:k-equal}), $\tau_{NL}^{cc}$ is 
\begin{eqnarray}
\tau_{NL}^{cc}(k,C_2,C_3,C_4)
\simeq 
-1.665\times10^{5} \times  \frac{\tilde \beta_4}{\alpha^{8/5}}
\left(\frac{({\cal P}_{\zeta}(k))^{1/2}}{4.8\times10^{-5}}\right)^{-8/5}
\frac{ \sum_{i=2,3,4} C_i^2}{ \sum_{i=2,3,4}\left(1+ C_i \right)^{-3/2}} .
\end{eqnarray}
We show the dependence of $\tau_{NL}^{cc}$ on the two angles $C_2$ and
$C_3$ in Fig.~\ref{fig:cc.eps}.

In the most symmetric case where $C_2=C_3=C_4=-1/3$,
$\tau_{NL}^{cc}$ becomes 
\begin{eqnarray}
\tau_{NL}^{cc}(k,C_2=-1/3,C_3=-1/3,C_4=-1/3)= -1.007 \times 10^4
\times  \frac{\tilde \beta_4}{\alpha^{8/5}}
\left(\frac{({\cal P}_{\zeta}(k))^{1/2}}{4.8\times10^{-5}}\right)^{-8/5}.
\end{eqnarray}

\subsection{scalar exchange contribution}

In this subsection we present the scalar exchange contribution to the
trispectrum. The scalar exchange contribution, corresponding to the
diagram shown in Fig.~\ref{fig:3-point.eps}, is defined as 
\begin{eqnarray}
&&\!\!\!\!\left\langle \pi_{\bf k_1}\pi_{\bf k_2}
\pi_{\bf k_3}\pi_{\bf k_4}\right\rangle_{se}
\nonumber\\
&&\equiv - \int^t_{-\infty} dt_2 \int^{t_2}_{-\infty} dt_1
\bigg\langle \biggl[ \int dx_1^3 \frac{\beta a(t_1)}{2M^2} 
\left( \frac{d}{d t_1}\pi(t_1,x_1) \right)\left( \vec{\nabla} \pi(t_1,x_1)\right)^2, 
\nonumber \\
&&\qquad\qquad\qquad\qquad\qquad\qquad
\Bigl[\int dx_2^3 \frac{\beta a(t_2)}{2M^2} 
\left( \frac{d}{d t_2}\pi(t_2,x_2) \right)\left( \vec{\nabla} \pi(t_2,x_2)\right)^2, 
\pi_{\bf k_1}\pi_{\bf k_2}\pi_{\bf k_3}\pi_{\bf k_4} \Bigr]\biggr] \bigg\rangle.
\end{eqnarray}
If the bispectrum (\ref{eqn:bispectrum-zeta}) from ghost inflation is
observed then it determines $\beta$ and thus the scalar exchange
contribution to the trispectrum. Then the full trispectrum from ghost
inflation, which is the sum of this contribution and the contact term
contribution described in the previous subsection, is fully specified by
the other parameter $\tilde{\gamma}$ (or equivalently $\gamma$).

\begin{figure}[t]
\begin{minipage}{0.42\textwidth}
  \begin{center}
    \includegraphics[keepaspectratio=true,height=50mm]{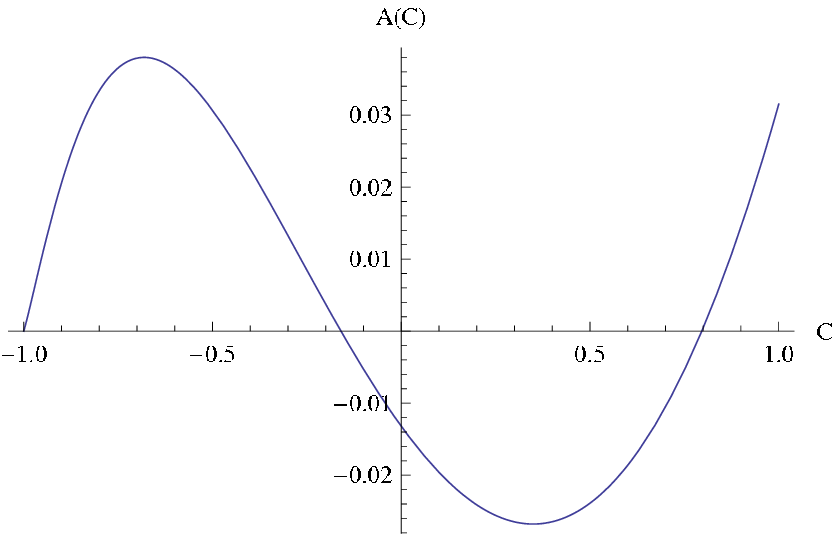}
  \end{center}
  \caption{This is the plot of the function $A(C)$.}
  \label{fig:A.eps}
  \end{minipage} 
  \begin{minipage}{0.05\textwidth} 
  \begin{center}
  \end{center}
  \end{minipage} 
  \begin{minipage}{0.42\textwidth}
  \begin{center}
    \includegraphics[keepaspectratio=true,height=50mm]{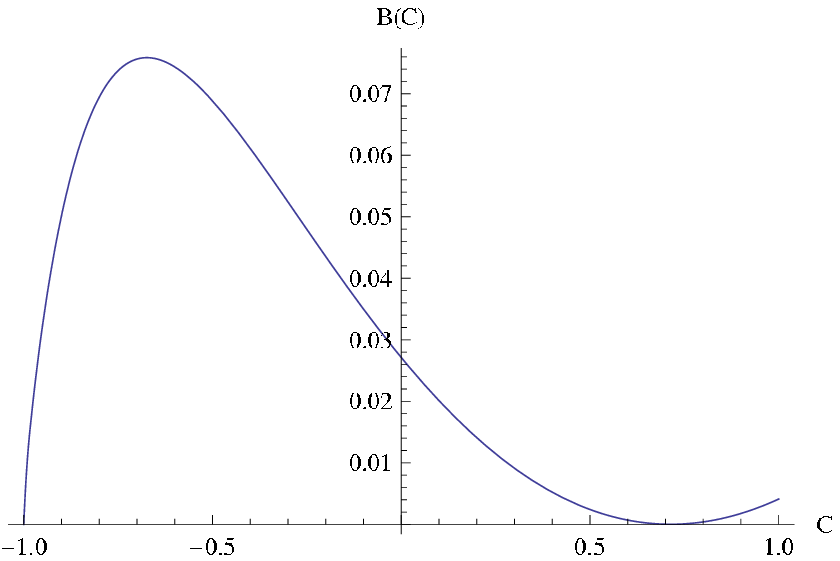}
  \end{center}
  \caption{This is the plot of the function $B(C)$.}
  \label{fig:B.eps}
\end{minipage}   
\end{figure}

Trispectrums $T_{\pi ,se}(\eta,{\bf k_1},{\bf k_2},{\bf k_3},{\bf k_4})$
and $T_{\zeta ,se}(\eta,{\bf k_1},{\bf k_2},{\bf k_3},{\bf k_4})$ are defined 
in the same manner as in the case of the contact term contribution, 
\begin{eqnarray}
&&\left\langle 
\pi_{\bf k_1}(t)\pi_{\bf k_2}(t)\pi_{\bf k_3}(t)\pi_{\bf k_4}(t)
\right\rangle _{se}  \equiv (2\pi)^3 \delta^3
({\bf k_1}+{\bf k_2}+{\bf k_3}+{\bf k_4}) 
T_{\pi ,se}(\eta,{\bf k_1},{\bf k_2},{\bf k_3},{\bf k_4}).
\\
&&\left\langle 
\zeta_{\bf k_1}(t)\zeta_{\bf k_2}(t)\zeta_{\bf k_3}(t)\zeta_{\bf k_4}(t)
\right\rangle _{se}  \equiv (2\pi)^3 \delta^3
({\bf k_1}+{\bf k_2}+{\bf k_3}+{\bf k_4}) 
T_{\zeta ,se}(\eta,{\bf k_1},{\bf k_2},{\bf k_3},{\bf k_4}),
\end{eqnarray}
and are related to each other as
\begin{eqnarray}
T_{\zeta ,se}(\eta,{\bf k_1},{\bf k_2},{\bf k_3},{\bf k_4})
=\left(\frac{H}{M^2} \right)^4 
T_{\pi ,se}(\eta,{\bf k_1},{\bf k_2},{\bf k_3},{\bf k_4}).
\end{eqnarray}
As in the calculations of other quantities, we take the limit 
$\eta_e\to 0$. Then, as calculated in Appendix~\ref{calculations} and
presented in (\ref{eq:Tpise}), we obtain 
\begin{eqnarray} 
&&T_{\zeta ,se}(\eta=0,{\bf k_1},{\bf k_2},{\bf k_3},{\bf k_4})
\nonumber\\
&&= 
\frac{\beta_3^2 \alpha  }{4} 
\left(\frac{H}{M}\right)^{14}\left(\frac{\pi}{8}\right)^3(2\pi)^3 
\delta^3({\bf k_1}+{\bf k_2}+{\bf k_3}+{\bf k_4}) 
 \biggl\{  -
\int^{\eta}_{-\infty}d\eta_2\int^{\eta_2}_{-\infty}d\eta_1
\nonumber \\ 
&& \quad
\Bigl(   \bigl\{ \bigl[
({\bf k_1} \cdot {\bf k_2}) (k_1^2 +k_2^2 +2 {\bf k_1} \cdot {\bf k_2})
\bigl( (-\eta_1)^{3/2} H_{3/4}^{(1)} (q_1 \eta_1^2) \bigr)
\bigl( (-\eta_1)^{3/2} H_{3/4}^{(1)} (q_2 \eta_1^2) \bigr)
\bigl( (-\eta_1)^{3/2} H_{-1/4}^{(1)} (q_{12} \eta_1^2) \bigr)
\nonumber \\
&& \qquad\qquad\quad
-
2(k_1^2 +{\bf k_1} \cdot {\bf k_2}) k_2^2
\bigl( (-\eta_1)^{3/2} H_{3/4}^{(1)} (q_1 \eta_1^2) \bigr)
\bigl( (-\eta_1)^{3/2} H_{-1/4}^{(1)} (q_2 \eta_1^2) \bigr)
\bigl( (-\eta_1)^{3/2} H_{3/4}^{(1)} (q_{12} \eta_1^2) \bigr)
\bigr]
\nonumber \\
&&\qquad\times 
\bigl[
({\bf k_3} \cdot {\bf k_4})(k_3^2 +k_4^2 +2 {\bf k_3} \cdot {\bf k_4})
\bigl( (-\eta_2)^{3/2} H_{3/4}^{(1)} (q_3 \eta_2^2) \bigr)
\bigl( (-\eta_2)^{3/2} H_{3/4}^{(1)} (q_4 \eta_2^2) \bigr)
\bigl( (-\eta_2)^{3/2} H_{-1/4}^{(2)} (q_{34} \eta_2^2) \bigr) 
\nonumber \\
&& \qquad\qquad\quad
-
2(k_3^2 +{\bf k_3} \cdot {\bf k_4}) k_4^2
\bigl( (-\eta_2)^{3/2} H_{3/4}^{(1)} (q_3 \eta_2^2) \bigr)
\bigl( (-\eta_2)^{3/2} H_{-1/4}^{(1)} (q_4 \eta_2^2) \bigr)
\bigl( (-\eta_2)^{3/2} H_{3/4}^{(2)} (q_{34} \eta_2^2) \bigr)
\bigr]
\nonumber \\
&& \qquad\qquad \times
\bigl(-\eta w_{-k_1}^*(\eta)\bigr)\bigl(-\eta w_{-k_2}^*(\eta)\bigr)
\bigl(-\eta w_{-k_3}^*(\eta)\bigr)
\bigl(-\eta w_{-k_4}^*(\eta)\bigr) 
\nonumber \\
&&\qquad\qquad\qquad\qquad\qquad\qquad\qquad\qquad
\qquad\qquad\qquad
+\mbox{23 permutational terms}\bigr\}+ c.c.\Bigr) 
\nonumber \\
&&
+\biggl[\biggl(\int^{\eta}_{-\infty} d\eta_1 \bigl[
({\bf k_1} \cdot {\bf k_2}) (k_1^2 +k_2^2 +2 {\bf k_1} \cdot {\bf k_2})
\bigl( (-\eta_1)^{3/2} H_{3/4}^{(1)} (q_1 \eta_1^2) \bigr)
\bigl( (-\eta_1)^{3/2} H_{3/4}^{(1)} (q_2 \eta_1^2) \bigr)
\bigl( (-\eta_1)^{3/2} H_{-1/4}^{(1)} (q_{12} \eta_1^2) \bigr)
\nonumber \\
&& \quad\qquad\qquad\qquad
-
2(k_1^2 +{\bf k_1} \cdot {\bf k_2}) k_2^2
\bigl( (-\eta_1)^{3/2} H_{3/4}^{(1)} (q_1 \eta_1^2) \bigr)
\bigl( (-\eta_1)^{3/2} H_{-1/4}^{(1)} (q_2 \eta_1^2) \bigr)
\bigl( (-\eta_1)^{3/2} H_{3/4}^{(1)} (q_{12} \eta_1^2) \bigr)
\bigr]
\biggr)
\nonumber\\
&& \times
\biggl(\int^{\eta}_{-\infty} d\eta_2 \bigl[
({\bf k_3} \cdot {\bf k_4})(k_3^2 +k_4^2 +2 {\bf k_3} \cdot {\bf k_4})
\bigl( (-\eta_2)^{3/2} H_{3/4}^{(2)} (q_3 \eta_2^2) \bigr)
\bigl( (-\eta_2)^{3/2} H_{3/4}^{(2)} (q_4 \eta_2^2) \bigr)
\bigl( (-\eta_2)^{3/2} H_{-1/4}^{(2)} (q_{34} \eta_2^2) \bigr) 
\nonumber \\
&& \quad\qquad\qquad\qquad
-
2(k_3^2 +{\bf k_3} \cdot {\bf k_4}) k_4^2
\bigl( (-\eta_2)^{3/2} H_{3/4}^{(2)} (q_3 \eta_2^2) \bigr)
\bigl( (-\eta_2)^{3/2} H_{-1/4}^{(2)} (q_4 \eta_2^2) \bigr)
\bigl( (-\eta_2)^{3/2} H_{3/4}^{(2)} (q_{34} \eta_2^2) \bigr)
\bigr]
\biggr)
\nonumber \\
&& \qquad\qquad\qquad\qquad\qquad\qquad\qquad\qquad\qquad\qquad \times
\bigl(-\eta w_{-k_1}^*(\eta)\bigr)\bigl(-\eta w_{-k_2}^*(\eta)\bigr)
\bigl(-\eta w_{k_3}(\eta)\bigr)
\bigl(-\eta w_{k_4}(\eta)\bigr) 
\nonumber \\
&&\qquad\qquad\qquad\qquad\qquad\qquad\qquad\qquad\qquad\qquad\qquad
\qquad\qquad\qquad\qquad\qquad\qquad
+\mbox{23 permutational terms} \biggr]\biggr\} ,
\end{eqnarray}
where $q_i$ is defined by eq.~(\ref{qi}) and 
\begin{eqnarray}
q_{ij}\equiv \frac{\sqrt{\alpha}H ({\bf k_i}+{\bf k_j})^2}{2M}.
\label{qij}
\end{eqnarray}

\begin{figure}[t]
\begin{minipage}{0.42\textwidth}
  \begin{center}
    \includegraphics[keepaspectratio=true,height=60mm]{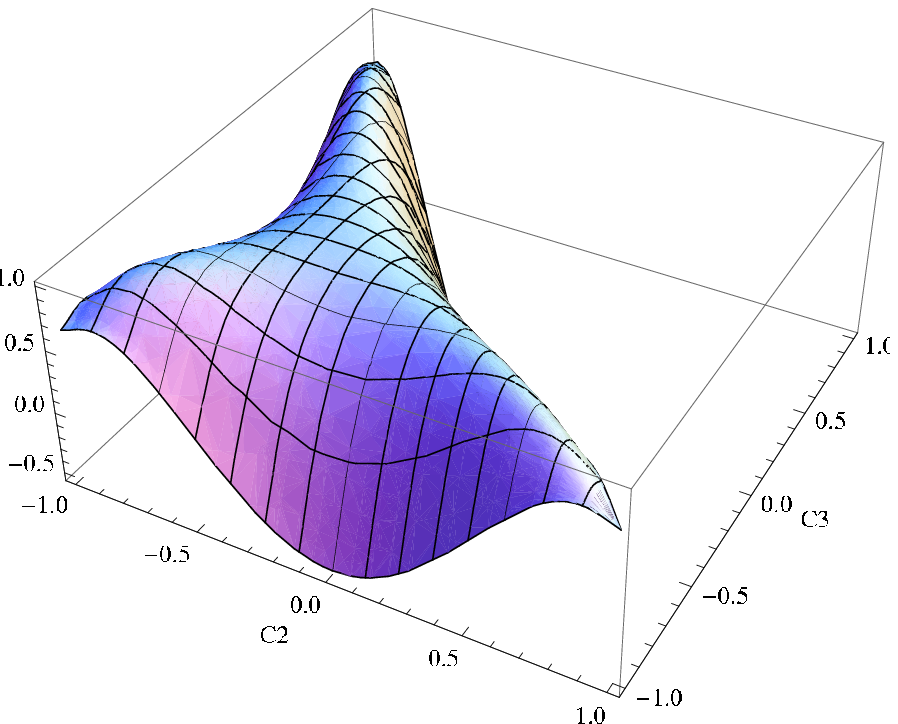}
  \end{center}
  \caption{The plot of the sum $A(C_2)+A(C_3)+A(C_4)$ as a
 function of $C_2$ and $C_3$, where $C_4=-1-C_2-C_3$. 
 The horizontal axis is rescaled by the value at $C_2=C_3=-1/3$.}
  \label{fig:f.eps}
  \end{minipage} 
  \begin{minipage}{0.05\textwidth} 
  \begin{center}
  \end{center}
  \end{minipage} 
  \begin{minipage}{0.42\textwidth}
  \begin{center}
    \includegraphics[keepaspectratio=true,height=60mm]{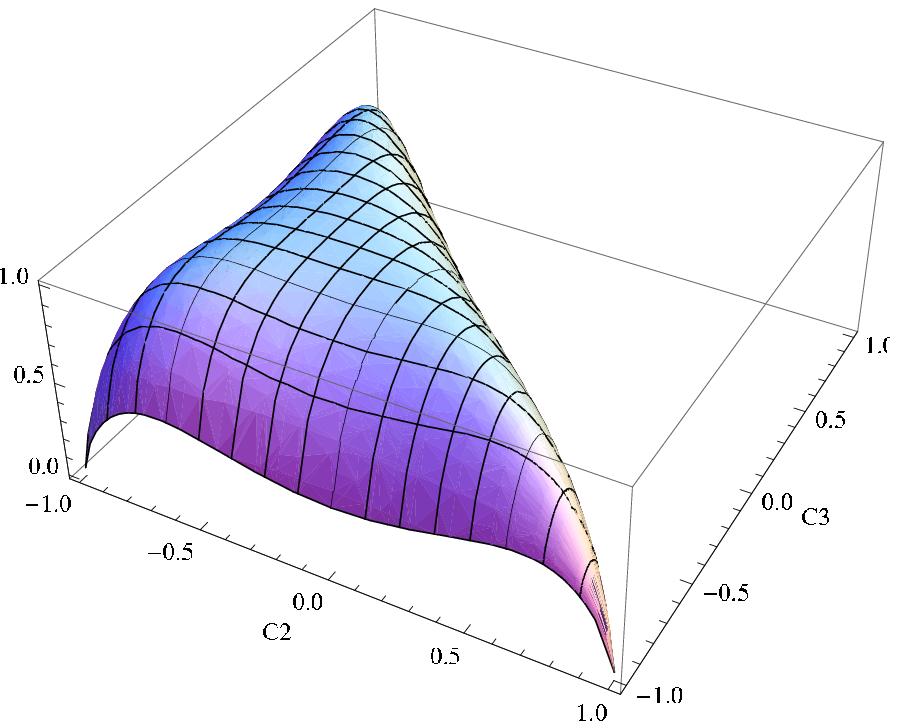}
  \end{center}
  \caption{The plot of the sum $B(C_2)+B(C_3)+B(C_4)$ as a
 function of $C_2$ and $C_3$ where $C_4=-1-C_2-C_3$. 
 The horizontal axis is rescaled by the value at $C_2=C_3=-1/3$.}
  \label{fig:g.eps}
\end{minipage}   
\end{figure}

The scalar exchange contribution to the trispectrum 
$T_{\zeta ,se}(\eta=0,{\bf k_1},{\bf k_2},{\bf k_3},{\bf k_4})$ 
depends on six parameters. As in the case of the contact term contribution, 
we consider the equilateral configurations by setting the amplitudes of all
momenta to be the same (see eq.(\ref{eq:k-equal})). Then the remaining
independent variables are ${\bf k_1}\cdot {\bf k_2}/(k_1k_2)$ 
and ${\bf k_1}\cdot {\bf k_3}/(k_1k_3)$. As shown in
eq.(\ref{Tzetase}), for equilateral configurations the scalar exchange
contribution is reduced to 
\begin{eqnarray}
&&T_{\zeta,se}(k,C_2,C_3,C_4)=
1.190 \times 10^{-16} 
\left(\frac{\bigl({\cal P}_\zeta (k)\bigr)^{1/2}}{4.8\times 10^{-5}}\right)^{22/5} 
\frac{\beta^2}{\alpha^{8/5}}
k^{-9} \sum_{i=2,3,4}
\left[A(C_i)+B(C_i)\right],\\
&&A(C)= \sqrt{2}\pi (1+C)^2  
\int^{\infty}_{0} dy_2 \int^{\infty}_{y_2} dy_1 y_1^{9/2} y_2^{9/2} 
\nonumber\\ 
&& \qquad\qquad\times
\Biggl\{
C \Bigl( K_{3/4}(y_1^2) \Bigr)^2
 K_{-1/4}\left(2(1+C)y_1^2\right) 
 - K_{3/4}(y_1^2)K_{-1/4}(y_1^2) K_{3/4}\left(2(1+C)y_1^2\right) 
\Biggr\}
\nonumber\\
&&\qquad\qquad\times
\Biggl\{
C \Bigl( K_{3/4}(y_2^2) \Bigr)^2
 I_{-1/4}\left(2(1+C)y_2^2\right) 
 + K_{3/4}(y_2^2)K_{-1/4}(y_2^2) I_{3/4}\left(2(1+C)y_2^2\right) 
\Biggr\} \label{A_i}\\
&&B(C) = (1+C)^2
\nonumber\\ 
&& \qquad\qquad\times
\Biggl| 
\int^{0}_{\infty}dy \  y^{9/2} \Biggl\{C \Bigl( K_{3/4}(y^2) \Bigr)^2
 K_{-1/4}\left(2(1+C)y^2\right) 
 - K_{3/4}(y^2)K_{-1/4}(y^2) K_{3/4}\left(2(1+C)y^2\right) 
\Biggr\}
 \Biggr|^2 ,
 \nonumber\\
 && \ \label{B_i}
\end{eqnarray}
where $C_i$ ($i=2,3,4$) are defined by eqs.~(\ref{eq:C_i}) and
constrained by momentum conservation as (\ref{momcon}). 
The scalar exchange contribution $T_{\zeta,se}(k,C_2,C_3,C_4)$ is
decomposed into six parts: those represented by $A(C_i)$ and $B(C_i)$
($i=2,3,4$). The functions $A(C)$ and $B(C)$ defined above are shown in
Fig~\ref{fig:A.eps} and Fig~\ref{fig:B.eps}, respectively. 
Fig~\ref{fig:f.eps} and Fig~\ref{fig:g.eps} show the sums
$A(C_2)+A(C_3)+A(C_4)$ and $B(C_2)+B(C_3)+C(C_4)$, respectively, with
$C_4=-1-C_2-C_3$ as functions of $C_2$ and $C_3$.

\begin{figure}[t]
\begin{minipage}{0.42\textwidth}
  \begin{center}
    \includegraphics[keepaspectratio=true,height=60mm]{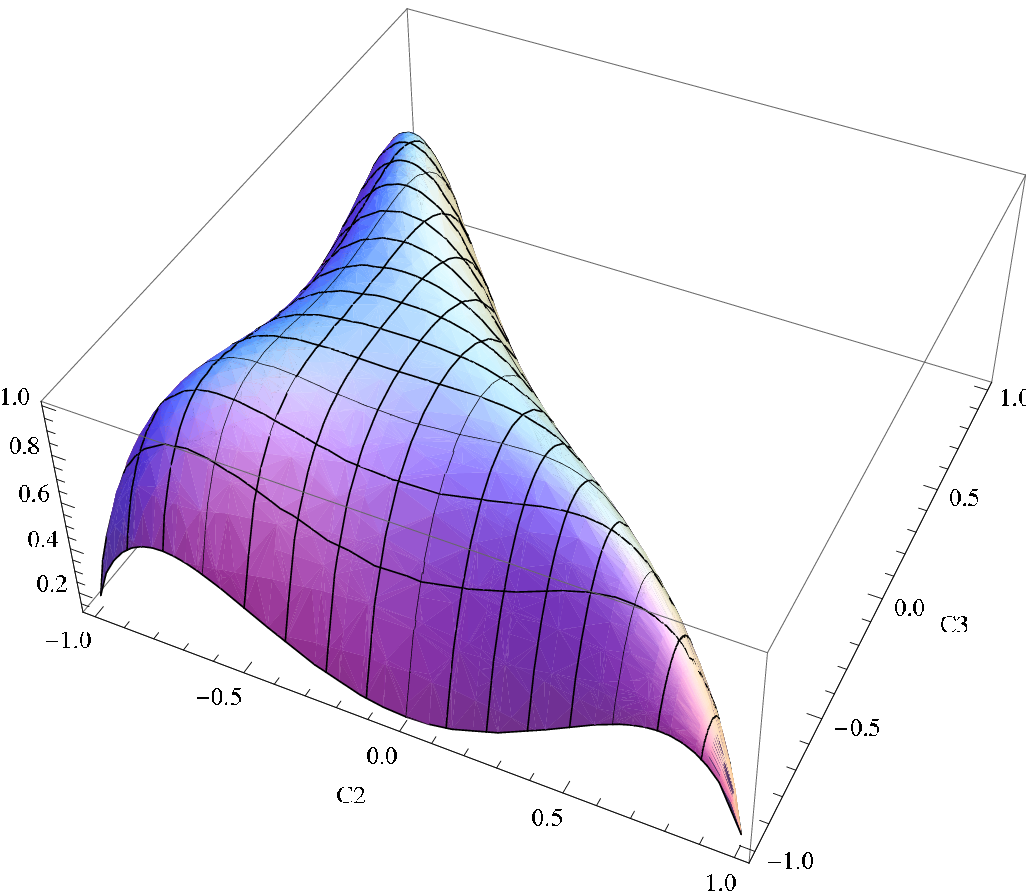}
  \end{center}
  \caption{
 This plot shows the dependence of 
 $T_{\zeta,se}(k,C_2,C_3,C_4)$ on $C_2$ and $C_3$, where
 $C_4=-1-C_2-C_3$. 
 The horizontal axis is rescaled by the value at $C_2=C_3=-1/3$.}
  \label{fig:seT.eps}
  \end{minipage} 
  \begin{minipage}{0.05\textwidth} 
  \begin{center}
  \end{center}
  \end{minipage} 
  \begin{minipage}{0.42\textwidth}
  \begin{center}
    \includegraphics[keepaspectratio=true,height=60mm]{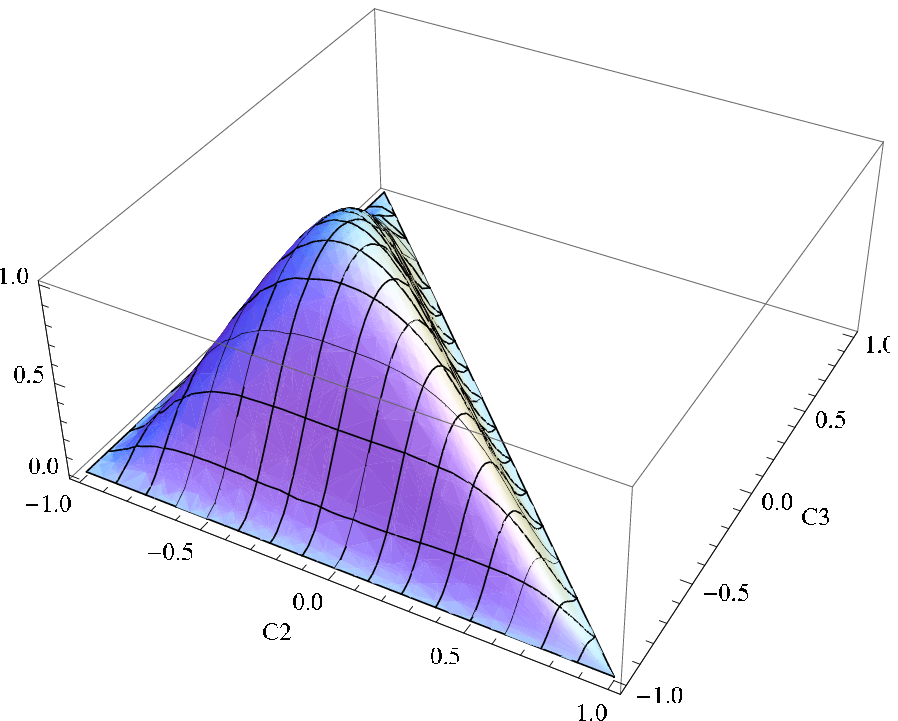}
  \end{center}
  \caption{
   This plot shows the dependence of 
   $\tau_{NL}^{se}(k,C_2,C_3,C_4)$ on $C_2$ and $C_3$, where
   $C_4=-1-C_2-C_3$. 
   The horizontal axis is rescaled by the value at $C_2=C_3=-1/3$.}
  \label{fig:se.eps}
\end{minipage}
\end{figure}

Fig.~\ref{fig:seT.eps} is the plot of the total scalar exchange
contribution $T_{\zeta,se}(k,C_2,C_3,C_4)$, which is the sum of the six 
parts. In the most symmetric case where $C_2=C_3=C_4=-1/3$, 
the trispectrum has the largest value
\begin{eqnarray}
T_{\zeta , se}(k,C_2=-\frac{1}{3},C_3=-\frac{1}{3},C_4=-\frac{1}{3})
= 2.561 \times 10^{-17} \left(
\frac{\bigl( {\cal P}_\zeta(k)\bigr)^{1/2}}{4.8 \times 10^{-5}}\right)^{22/5}
\frac{\beta^2}{\alpha ^{8/5}} k^{-9}.
\end{eqnarray}
This is always positive. Actually, Fig.~\ref{fig:seT.eps} shows that the
scalar exchange contribution to the equilateral trispectrum
$T_{\zeta,se}(k,C_2,C_3,C_4)$ is positive in the all parameter region.

Following Ref.~\cite{Arroja:2009pd}, we define the nonlinear parameter
$\tau^{se}_{NL}$ as 
\begin{eqnarray}
\tau_{NL}^{se}({\bf k_1},{\bf k_2},{\bf k_3},{\bf k_4})&\!\! \equiv\!\!& 
\frac{T_{\zeta, se}({\bf k_1},{\bf k_2},{\bf k_3},{\bf k_4})}
{(2 \pi^2 {{\cal P_{\zeta}}( k)})^3}
\prod_{i=1}^4 k_i^3
\nonumber\\
&&\times \left[ 
\left(k_1^3 k_2^3+k_3^3 k_4^3\right)\left(k_{13}^{-3}+k_{14}^{-3} \right) 
+\left(k_1^3 k_4^3+k_2^3 k_3^3\right)\left(k_{12}^{-3}+k_{13}^{-3} \right) 
+\left(k_1^3 k_3^3+k_2^3 k_4^3\right)\left(k_{12}^{-3}+k_{14}^{-3} \right) 
\right]^{-1}. \nonumber \\
&& \qquad
\end{eqnarray}
As shown in eq.~(\ref{tause}), $\tau^{se}_{NL}$ for the equilateral
configurations is 
\begin{eqnarray}
\tau_{NL}^{se}(k,C_2,C_3,C_4)\simeq
8.945 \times10^5\times
\frac{\beta_3^2}{\alpha^{8/5}} 
\left(\frac{({\cal P}_{\zeta}(k))^{1/2}}{4.8\times10^{-5}}\right)^{-8/5}
\frac{\sum_{i=2,3,4}\left[A(C_i)+B(C_i)\right]}
{\sum_{i=2,3,4}(1+C_i)^{-3/2}}.
\end{eqnarray}
Fig.~\ref{fig:se.eps} shows the plot of the shape of 
$\tau^{se}_{NL}$.

In the most symmetric case where $C_2=C_3=C_4=-1/3$,
$\tau_{NL}^{se}$ becomes 
\begin{eqnarray}
\tau_{NL}^{se}(k,C_2=-1/3,C_3=-1/3,C_4=-1/3)\simeq 3.494 \times 10^4
\times
\frac{\beta_3^2}{\alpha^{8/5}} 
\left(\frac{({\cal P}_{\zeta}(k))^2}{4.8\times10^{-5}}\right)^{-8/5}.
\end{eqnarray}

\section{Summary and Discussion}
\label{Sum}

In this work we have calculated and investigated the trispectrum of
curvature perturbation generated during ghost inflation. 

The analysis of scaling dimensions of operators makes it possible to
identify the leading diagrams contributing to the trispectrum in ghost
inflation. Actually, there are two leading-order contributions. One is
represented by a diagram with one four-point vertex. This contribution is
called contact term contribution. The other is represented by a diagram
with two three-point vertices and called scalar exchange
contribution. We have analyzed these two contributions separately.

We have obtained general expressions for the two contributions as
functions of six independent parameters. The six parameters are
amplitudes of four $3$-momenta and two angles between momenta. (Note
that the sum of four $3$-momenta must vanish because of the momentum
conservation.)

In order to calculate the concrete values, we have focused on the
equilateral case where all momenta has the same amplitude and where
there remain two independent angular parameters as well as an overall
amplitude of $3$-momenta. Then we have calculated the non-linear 
parameters $\tau_{NL}^{cc}$ and $\tau_{NL}^{se}$ for the contact term
contribution and the scalar exchange contribution, respectively.

In the case of local-type non-Gaussianity, it was forecasted that
PLANCK will give the constraint 
$|\tau_{NL}|\sim 560$~\cite{Kogo:2006kh}. 
In the present paper we have shown that $\tau_{NL}^{cc}$ and
$\tau_{NL}^{se}$ are typically of order $O(10^4)$. Therefore, the
trispectrum from ghost inflation is probably detectable by PLANCK. 
Note, however, that the meaning of the non-linear parameter $\tau_{NL}$
is different for different types of non-Gaussianities. (The same is true
for the non-linear parameter $f_{NL}$ of bispectra.) Therefore, as a
future work, it is important to investigate detectability of the 
trispectrum predicted by ghost inflation in more detail.

Now let us compare our results with the trispectrum from DBI
inflation calculated in Refs.~\cite{Arroja:2009pd,Mizuno:2009mv}. 
The overall behaviors of trispectra are indeed similar. The
trispectrum from ghost inflation has a peak at the equilateral
configurations, i.e. when all four $3$-momenta have the same amplitude,
because non-Gaussianity is mainly generated in the horizon-crossing
epoch. This feature of trispectrum is shared with DBI
inflation. Moreover, the dependence of the equilateral trispectrum on
the angular variables $C_2$ and $C_3$ also has similarities. 
In both DBI inflation and ghost
inflation, the scalar exchange contribution has the maximum value at the
most symmetric point $C_2=C_3=C_4=-1/3$, and the absolute value of
the contact term contribution becomes minimum at that point. 
The contact term contribution to the equilateral trispectrum has similar
dependence on $C_2$ and $C_3$ in the two models of inflation. This can
be easily seen by comparing Fig.~\ref{fig:Tcc.eps} in the present paper
and the right figure of Fig.1 in Ref.~\cite{Mizuno:2009mv}.

There are also some differences between the trispectrum from DBI 
inflation and that from ghost inflation. In DBI inflation, the value of 
the equilateral trispectrum is almost constant except for the edge
region near the boundaries defined by $C_i=-1$ ($i=2,3,4$). This feature
can be seen in, e.g., the left figure of Fig.1 of
Ref.~\cite{Mizuno:2009mv}. The trispectrum rapidly decreases near the
boundaries $C_i=-1$ and the plateau looks like a triangle. On the other
hand, as we can see in Fig.\ref{fig:seT.eps} of the present paper, in
ghost inflation the value of the equilateral trispectrum smoothly
decreases towards the boundaries $C_i=-1$ ($i=2,3,4$). As a result the
shape of the plateau looks different from that in DBI inflation.

If we look into actual values, we find an important difference. 
(Note that those figures mentioned above are normalized by the values at 
$C_2=C_3=C_4=-1/3$ and thus do not tell the actual values of
trispectra.) The sign and magnitude of the contact term contribution depend
on the sign and magnitude of the quartic coupling constant. (This is in
contrast to the scalar exchange contribution, whose sign does not depend
on the sign of the cubic coupling constant.) In DBI inflation, the
quartic coupling constant is determined by the sound speed. As a result,
the equilateral trispectrum from DBI inflation has a positive value at
the most symmetric configuration $C_2=C_3=C_4=-1/3$. On the other hand,
in ghost inflation the dimensionless quartic coupling constant is an arbitrary
parameter of order unity and thus the equilateral trispectrum at
$C_2=C_3=C_4=-1/3$ can be either positive or negative.  Therefore, if
equilateral-type non-Gaussianity is detected either by bispectrum or by
trispectrum and if the negative equilateral trispectrum at the most
symmetric point is observed, then it will support the ghost inflation
scenario.

\vspace{1cm}
\textit{Note added}: 
While we were preparing the present paper, ref.~\cite{Huang:2010ab}
appeared on the arXiv. Before that, we had finished all calculations and
presented our results at the workshop  "The non-Gaussian universe" at 
Yukawa Institute for Theoretical Physics on March 25, 2010. The
presentation file has been available from the workshop website at 
\verb+http://www2.yukawa.kyoto-u.ac.jp/~nlg/2010_3/program.htm+ 
since March 27, 2010.

\section*{Acknowledgement}
K.~I. thanks Masato Minamitsuji for many useful discussions. The authors
are grateful to the organizers of the workshop  "The non-Gaussian
universe" at Yukawa Institute for Theoretical Physics for their warm 
hospitality and providing stimulating environment during the
workshop. 
The work of S.M. is supported by JSPS Grant-in-Aid for Young Scientists
(B) No.~17740134, JSPS Grant-in-Aid for Creative Scientific Research
No.~19GS0219, MEXT Grant-in-Aid for Scientific Research on Innovative
Areas No.~21111006, JSPS Grant-in-Aid for Scientific Research (C)
No.~21540278, and the Mitsubishi Foundation. This work was supported by
World Premier International Research Center Initiative.

\appendix
\section{Review of in-in formalism}
\label{in-in}

In Heisenberg picture, the expectation value of an observable is
\begin{eqnarray}
\left\langle\Omega | Q(t) | \Omega \right\rangle,
\end{eqnarray}
where $Q(t)$ is a Heisenberg operator corresponding to the observable
and $\left| \Omega \right\rangle$ is the vacuum state in the initial
time. A Heisenberg operator is sandwiched between initial vacuum
states~\cite{Weinberg:2005vy}. 

The time evolution of the Heisenberg operator is 
\begin{eqnarray}
Q(t)=e^{iH(t-t_0)} Q(t_0) e^{-iH(t-t_0)},
\end{eqnarray}
where $H$ is Hamiltonian. 
On the other hand, an operator in interaction picture evolves 
by the Hamiltonian $H_{free}$ of the free theory as
\begin{eqnarray}
Q_I(t)=e^{iH_{free}(t-t_0)} Q(t_0) e^{-iH_{free}(t-t_0)}.
\end{eqnarray}
The relation between these operators is 
\begin{eqnarray}
Q(t)= \left[ U(t,t_0) \right]^\dagger Q_I(t) U(t,t_0),
\end{eqnarray}
where
\begin{eqnarray}
U(t,t') &\!\!\equiv\!\!& e^{iH_{free}(t-t')}e^{-iH_(t-t')} 
\nonumber\\
&\eq&
T\left\{\exp\left[-i \int_{t'}^t dt'' H_I(t'')
\right]
\right\} \qquad(t\ge t'),
\end{eqnarray}
\begin{eqnarray}
H_I=H-H_{free},
\end{eqnarray}
and $T$ indicates time-ordered products. 

Let us now consider the time evolution of the vacuum state. 
First, we evolve the vacuum state $\left|0\right\rangle_{free}$ 
of $H_{free}$ by the total Hamiltonian $H$: 
\begin{eqnarray}
e^{-iH(t_0-t_{in})} \left|0\right\rangle_{free}
&\eq& e^{-iE_0(t_0-t_{in})} \left|\Omega\right\rangle 
\left\langle\Omega|0\right\rangle_{free} +\sum_n
e^{-iE_n(t_0-t_{in})}\left| n\right\rangle 
\left\langle n|0\right\rangle_{free}
\nonumber\\
&\to& e^{-iE_0(t_0-t_{in})} \left|\Omega\right\rangle 
\qquad(t_{in}\to -\infty(1-i\epsilon)),
\label{eqn:iepsilon}
\end{eqnarray}
where $\left|n\right\rangle$ ($n=1,2,\cdots$) represent excited
eigenstates of $H$ and 
\begin{eqnarray}
H \left|\Omega\right\rangle=E_0 \left|\Omega\right\rangle, 
\qquad
H \left|n\right\rangle=E_n\left|n\right\rangle.
\end{eqnarray}
Thus,
\begin{eqnarray}
\left|\Omega\right\rangle= 
 C\, U(t_0,-\infty)\left| 0\right\rangle_{free},
\end{eqnarray}
where $C$ is a constant factor.
Similarly, we have 
\begin{eqnarray}
\left\langle\Omega\right|=
C^* \,_{free}\!\!\left\langle0\right| \left[U(t_0,-\infty)\right]^\dagger.
\end{eqnarray}
By taking the inner product $\left\langle\Omega | \Omega\right\rangle$,
we can determine $C C^*$ as 
\begin{eqnarray}
1=\left\langle\Omega | \Omega\right\rangle=C C^*.
\end{eqnarray}
Eventually, we have 
\begin{eqnarray}
\left\langle\Omega | Q(t) | \Omega\right\rangle
&\eq& \,_{free}\!\!\left\langle0\right| \left[U(t_0,-\infty)\right]^\dagger
Q(t) U(t_0,-\infty)\left| 0\right\rangle_{free}
\nonumber\\ 
&\eq& \,_{free}\!\!\left\langle0\right| \left[U(t,-\infty)\right]^\dagger
Q_I(t) U(t,-\infty)\left| 0\right\rangle_{free}
\nonumber\\
&\eq&\sum_{N=0}^\infty i^N \int_{-\infty}^t \!\! dt_N
\int_{-\infty}^{t_N} \!\! dt_{N-1} \cdots \int_{-\infty}^{t_2}\!\! dt_1 
\,_{free}\!\!\left\langle0\right|
 [H_I(t_1),[H_I(t_2),\cdots[H_I(t_N),Q_I(t)]\cdots]]
\left| 0\right\rangle_{free}.
\end{eqnarray}

\section{Interaction term of Hamiltonian}
\label{interactionhamiltonian}

If time derivatives of fields are included in interaction terms in a
Lagrangian, the interaction terms in the Hamiltonian are not
simply the minus the corresponding terms in the Lagrangian. The time
evolution of a quantum state is driven by a Hamiltonian (See
Appendix~\ref{in-in}) but a theory is usually defined in terms of a
Lagrangian. Thus, it is important to understand the difference between 
interaction terms in a Lagrangian and those in a Hamiltonian. In
this appendix, we consider this issue and obtain a perturbative
expression for the difference. In Ref~\cite{Huang:2006eha}, this problem was
investigated for a system with one particle. This appendix extends it to
a multi particle system.

We consider the case where the Lagrangian is of the following form: 
\begin{eqnarray}
&&L(q^a, \dot q^a,t) = L_{free} +L_{int}(q^a,\dot q^a,t) ,\\
&&L_{free} \equiv  \frac{1}{2}g_{ab}(t)\dot q^a \dot q^b  
-\frac{1}{2} h_{ab}(t) q^a q^b 
+ A_{ab}(t) \dot q^a q^b,
\end{eqnarray}
where $g_{ab}$, $h_{ab}$ and $A_{ab}$ are independent of $q^a$ 
(but can depend on time) and 
$L_{int}(q^a,\dot q^a,t)$ represents terms of third or higher order 
in $q^a$ and $\dot q^a$. 
For simplicity, we consider the case where there is no constraint.
In this case the inverse matrix of $g_{ab}(t)$ exists and we denote it
by $g^{ab}(t)$. The canonical momenta are
\begin{eqnarray}
p_a \equiv \frac{\partial L}{\partial \dot q^a}
= g_{ab}(t)\dot q^b +A_{ab}(t)q^b + 
\frac{\partial L_{int}}{\partial \dot q^a} (q^a,\dot q^a (q^b,p_b,t),t) ,
\label{eq;p_a}
\end{eqnarray}
and the Hamiltonian is
\begin{eqnarray}
H(t)&\!\!\equiv \!\!& p_a \dot q^a (q^c, p_c,t) 
-L(q^a, \dot q^a (q^c, p_c,t),t) \nonumber\\ 
&\eq& p_a \dot q^a (q^c, p_c,t)
-\frac{1}{2}g_{ab}(t)\dot q^a (q^c, p_c,t) \dot q^b (q^c, p_c,t) 
+\frac{1}{2} h_{ab}(t) q^a q^b
-L_{int}(q^a, \dot q^a (q^c, p_c,t),t) 
\nonumber\\
&\eq& \frac{1}{2} g^{ab}(t) (p_a -A_{ac}(t)q^c)( p_b-A_{bd}(t)q^d) 
+\frac{1}{2} h_{ab}(t) q^a q^b 
-L_{int} (q^a,\dot q^a (q^b,p_b,t),t) 
\nonumber \\
&&\qquad\qquad\qquad\qquad
- \frac{1}{2} g^{ab} (t)
\frac{\partial L_{int}}{\partial \dot q^a}(q^c,\dot q^c (q^d,p_d,t),t)
\frac{\partial L_{int}}{\partial \dot q^b} (q^c,\dot q^c (q^d,p_d,t),t),
\label{Hintttt}
\end{eqnarray}
where $\dot q^a(q^b,p_b,t)$ is defined by solving eq.~(\ref{eq;p_a})
with respect to $\dot q^a$. 

Now, we quantize the theory. 
What we have to do is to promote the variables $q^a$ and $p_a$ to 
the corresponding operators $Q^a$ and $P_a$. 
Then we have the standard commutation relations, 
\begin{eqnarray}
&&[Q^a, P_b] = i \delta_b^{\ a}, \qquad [Q^a,Q^b] =[P_a,P_b] =0.\label{QP1}
\end{eqnarray}
A complete set of quantum states is given by the eigenstates of $Q^a$ or
$P_a$ which satisfy, respectively, 
\begin{eqnarray}
Q^a \left| q \right\rangle =q \left| q \right\rangle , \qquad
P_a \left| p \right\rangle =p \left| p \right\rangle ,
\end{eqnarray}
and the inner products of them are 
\begin{eqnarray}
\left\langle q' | q \right\rangle= \prod_a  \delta_{{q'}^a q^a},
\qquad
\left\langle p' | p \right\rangle= \prod_a  \delta_{{p'}_a p_a},
\qquad
\left\langle q | p \right\rangle= \prod_a \frac{1}{\sqrt{2 \pi}} 
\exp(iq^a p_a).\label{<qp>}
\end{eqnarray}

In order to use the in-in formalism reviewed in Appendix~\ref{in-in}, we
should adopt the interaction representation. The time-evolution equation
of the operators $Q^a_I$ and $P_{I,a}$ is a linear differential
equation, where the subscript ``$I$'' implies that these operators are
in the interaction representation. In the interaction representation, an
operator $O_I$ depends on time as 
\begin{eqnarray}
O_I(t) = \left[\bar T\left\{ \exp (i\int_{t_0}^t H_{free}(t') dt')
\right\}\right] O(t_0) 
\left[T\left\{\exp (-i \int_{t_0}^t H_{free}(t')dt')\right\}\right] ,
\end{eqnarray}
where 
\begin{eqnarray}
H_{free } (t)
\equiv  \frac{1}{2} g^{ab}(t) (P_a-A_{ac}(t)Q^c) (P_b- A_{bd}(t)Q^d) 
+\frac{1}{2} h_{ab}(t) Q^a Q^b .
\end{eqnarray}
Thus, the time derivative of the operator $O_I(t)$ is 
\begin{eqnarray}
\dot O_I(t) &\eq& i \left[\bar T\left\{ 
\exp (i\int_{t_0}^t H_{free}(t') dt')\right\}\right] H_{free}(t) O(t_0) 
\left[ T\left\{\exp (-i \int_{t_0}^t H_{free}(t')dt,)\right\}\right] 
 \nonumber\\
&&\qquad
-i \left[\bar T \left\{ \exp (i\int_{t_0}^t H_{free}(t') dt')
\right\}\right]  O(t_0) H_{free}(t) 
\left[T\left\{\exp (-i \int_{t_0}^t H_{free}(t')dt')\right\}\right]
 \nonumber\\
&\eq& -i [O_I(t),H_{free,I}(t)]
\end{eqnarray}

The commutation relations (\ref{QP1}) lead to the same commutation
relations for $Q^a_I$ and $P_{I,a}$ as 
\begin{eqnarray}
[Q_I^a(t), P_{I,b}(t)] = i \delta_b^{\ a}, \label{QP}
\qquad
[Q^a_I(t),Q^b_I(t)] =[P_{I,a}(t),P_{I,b}(t)] =0,
\end{eqnarray}
and their time derivatives are
\begin{eqnarray}
&&\dot P_{I,a}(t) = -h_{ab} (t)Q^b_I(t),\\
&&\dot Q^a_I (t)=g^{ab} (t)(P_{I,b}(t)-A_{bc}(t)Q^c_I(t)). \label{QaI}
\end{eqnarray}
Since these equations are the same as those for free particles, 
we can easily represent $Q^a_I$ and $P_{I,a}$ by using the creation and annihilation operators as in the case of free particles.

The time evolution of a state $\left| \psi(t) \right\rangle_I$ 
is represented as 
\begin{eqnarray}
i\frac{d}{dt} \left| \psi(t) \right\rangle_I =
H_{int,I}(t) \left| \psi(t) \right\rangle_I,
\label{evolupsi}
\end{eqnarray}
where 
\begin{eqnarray}
H_{int,I}(t) &\!\! \equiv\!\! & 
\left[\bar T\left\{ \exp (i\int_{t_0}^t H_{free}(t') dt')
\right\}\right]  H_{int}(Q^a,P_a,t) 
\left[ T\left\{ \exp (-i\int_{t_0}^t H_{free}(t') dt')
\right\}\right] 
\nonumber\\
&\eq& H_{int}(Q^a_I(t),P_{I,a}(t),t),
\end{eqnarray}
and 
\begin{eqnarray}
H_{int}(Q^a,P_a,t)&\eq&H(Q^a,P_a,t)-H_{free}(Q^a,P_a,t) \nonumber \\
&\eq&-L_{int} (Q^a,\dot q^a (Q^b,P_b),t) \nonumber \\
&&\qquad
- \frac{1}{2} g^{ab} (t)
\frac{\partial L_{int}}{\partial \dot q^a}(Q^c,\dot q^c (Q^d,P_d),t)
\frac{\partial L_{int}}{\partial \dot q^b} (Q^c,\dot q^c (Q^d,P_d),t).
\label{eq:intHam}
\end{eqnarray}

Solving eq.~(\ref{eq;p_a}) with respect to $\dot q^a$ iteratively, we
obtain 
\begin{eqnarray}
\dot q^a(Q^b,P_b,t) &\eq & 
g^{ab}(t) \left((P_b-A_{bc}(t)Q^c)-
\frac{\partial L_{int}}{\partial \dot q^b}(Q^c,\dot q^c(Q^c,P_c,t),t)
 \right) \nonumber\\ 
&\eq&
\dot Q^a - g^{ab}\frac{\partial L_{int}}{\partial \dot q^b}(Q^c,\dot Q^c,t)
+(\mbox{more than third order in $Q^a$ and $P_a$}), 
\end{eqnarray}
where we have used eq.~(\ref{QaI}).
Substituting this to eq.~(\ref{eq:intHam}), we can see the difference
between interaction terms in the Hamiltonian and those in the
Lagrangian: 
\begin{eqnarray}
&&H_{int}(Q^a_I,P_{I,a},t)
=-L_{int} (Q^a_I,\dot Q^a_I ,t)
+ \frac{1}{2} g^{ab} (t)
\frac{\partial L_{int}}{\partial \dot q^a}(Q^c_I,\dot Q^c_I,t)
\frac{\partial L_{int}}{\partial \dot q^b} (Q^c_I,\dot Q^c_I,t)
\nonumber\\ 
&&\qquad\qquad\qquad\qquad\qquad\qquad\qquad\qquad\qquad\qquad\qquad
+(\mbox{more than fifth order in $Q^a$ and $P_a$}).
\end{eqnarray}

\section{Detailed calculations}
\label{calculations}
In this appendix we will show the concrete calculations 
of the trispectrum from ghost inflation. 

\subsection{contact term contribution}

The contact term contribution to the four-point function can be written
as
\begin{eqnarray}
\left\langle 
\pi_{\bf k_1}(t)\pi_{\bf k_2}(t)\pi_{\bf k_3}(t)\pi_{\bf k_4}(t)
\right\rangle _{cc}
\equiv
i \int_{-\infty}^t dt' \left\langle
\left[\frac{\tilde \gamma}{8M^4} \int d^3x' \frac{ (\nabla \pi(t',x'))^4}{a(t')},
\pi_{\bf k_1}(t)\pi_{\bf k_2}(t)\pi_{\bf k_3}(t)\pi_{\bf k_4}(t)\right]
\right\rangle .
\end{eqnarray}
Transforming variables as
\begin{eqnarray}
  dt'=a(\eta')d\eta, \qquad 
\pi(t')\equiv \frac{u(\eta')}{a(\eta')} ,\qquad
a(\eta)=- \frac{1}{H\eta}  ,
\label{eq:transform}
\end{eqnarray}
we obtain
\begin{eqnarray}
&&\left\langle 
\pi_{\bf k_1}(t)\pi_{\bf k_2}(t)\pi_{\bf k_3}(t)\pi_{\bf k_4}(t)
\right\rangle _{cc} \nonumber\\
&&\qquad =
i\int_{-\infty}^\eta d\eta' \frac{\tilde \gamma H^8}{8M^4}
\left\langle\left[
\int d^3x' (-\nabla \eta' u(\eta',x'))^4, (-\eta u_{\bf k_1}(\eta))
(-\eta u_{\bf k_2}(\eta))(-\eta u_{\bf k_3}(\eta))(-\eta u_{\bf k_4}(\eta)) 
\right]\right\rangle \nonumber \\
&&\qquad = 
i \frac{\tilde \gamma H^8}{8 M^4} \int^\eta_{-\infty}  d\eta'
\int d^3 k'_1d^3 k'_2d^3 k'_3 ( {\bf k'_1} \cdot {\bf k'_2})
(-{\bf k'_1} \cdot {\bf k'_3} -{\bf k'_2}\cdot {\bf k'_3} -{k'_3}^2)
\nonumber \\
&&\qquad\qquad\qquad
\Bigl\langle\bigl [
(-\eta' u_{\bf k'_1}(\eta'))(-\eta' u_{\bf k'_2}(\eta'))
(-\eta' u_{\bf k'_3}(\eta'))
(-\eta' u_{-{\bf k'_1}-{\bf k'_2}-{\bf k'_3}}(\eta')) ,
\nonumber \\
&&\qquad\qquad\qquad\qquad\qquad\qquad\qquad\qquad\qquad
(-\eta u_{\bf k_1}(\eta))(-\eta u_{\bf k_2}(\eta))
(-\eta u_{\bf k_3}(\eta))(-\eta u_{\bf k_4}(\eta))
\bigr]\Bigr\rangle ,
\label{eq:4point}
\end{eqnarray}
where $u_{\bf k}(\eta)$ is written in terms of mode functions
$w_k(\eta)$, $w_{-k}^*(\eta)$ and operators 
$a_{\bf k}$ and $a_{-{\bf k}}^{\dagger}$ as 
\begin{eqnarray}
u_{\bf k}(\eta)=w_k(\eta) a_{\bf k} + w_{-k}^*(\eta ) a_{-{\bf k}}^\dagger,
\label{eq:u_k}
\end{eqnarray}
\begin{eqnarray}
[a_{\bf k},a_{{\bf k}'}^\dagger]=(2\pi)^3\delta^3({\bf k}-{\bf k}') .
\label{eq:exchange}
\end{eqnarray}
%
Substituting eq.~(\ref{eq:u_k}) into eq.~(\ref{eq:4point}),
we have
\begin{eqnarray}
&&\left\langle 
\pi_{\bf k_1}(t)\pi_{\bf k_2}(t)\pi_{\bf k_3}(t)\pi_{\bf k_4}(t)
\right\rangle _{cc} \nonumber\\
&&\qquad =
i \frac{\tilde \gamma H^8}{8 M^4} \int^\eta_{-\infty}  d\eta'
(2\pi)^3 \delta^3({\bf k_1}+{\bf k_2}+{\bf k_3}+{\bf k_4}) 
\bigl\{
({\bf  k_1} \cdot {\bf  k_2})({\bf  k_3} \cdot {\bf k_4}) +(\mbox{23 terms})
\bigr\}
\nonumber\\
&&\qquad\qquad\qquad\qquad\qquad\times
\bigl\{
(-\eta'w_{-k_1}(\eta'))(-\eta'w_{-k_2}(\eta'))
(-\eta'w_{-k_3}(\eta'))(-\eta'w_{-k_4}(\eta'))
\nonumber\\
&&\qquad\qquad\qquad\qquad\qquad\qquad\qquad
(-\eta w_{-k_1}^*(\eta))(-\eta w_{-k_2}^*(\eta))
(-\eta w_{-k_3}^*(\eta))(-\eta w_{-k_4}^*(\eta))
-\mbox{c.c} \bigr\} .
\label{eq:4point2}
\end{eqnarray}

The contact term contribution to the trispectrum 
$T^\eta_{\pi ,cc}({\bf k_1},{\bf k_2},{\bf k_3},{\bf k_4})$ is
\begin{eqnarray}
&&T^\eta_{\pi ,cc}({\bf k_1},{\bf k_2},{\bf k_3},{\bf k_4}) =
\frac{\tilde \gamma H^8}{8 M^4} 2\Re\Bigl\{
i \Bigl(
({\bf k_1} \cdot {\bf k_2})({\bf k_3} \cdot {\bf k_4}) +(\mbox{23 terms})
\Bigr)
\nonumber \\
&&\qquad\qquad\qquad\qquad\qquad\qquad\qquad
(-\eta w_{-k_1}^*(\eta))(-\eta w_{-k_2}^*(\eta))
(-\eta w_{-k_3}^*(\eta))(-\eta w_{-k_4}^*(\eta))
\nonumber\\
&&\qquad\qquad\qquad\qquad\qquad\qquad\qquad
\int^\eta_{-\infty} d \eta' 
(-\eta' w_{-k_1}(\eta'))(-\eta' w_{-k_2}(\eta'))
(-\eta' w_{-k_3}(\eta'))(-\eta' w_{-k_4}(\eta'))
\Bigr\}.
\end{eqnarray}

We are interested in the case where
the physical length scales of all arguments $k_i$ $(i=1,2,3,4)$ of the trispectrum 
are larger than the Hubble scale of de Sitter spacetime 
at the end of inflation, 
which means 
\begin{eqnarray}
k_i \eta \ll 1. 
\end{eqnarray}
Therefore, we approximate the trispectrum at the end of inflation
as $T^{\eta\to 0}_{\pi ,cc}({\bf k_1},{\bf k_2},{\bf k_3},{\bf k_4})$. 
Then, we have
\begin{eqnarray}
&&T_{\pi ,cc}({\bf k_1},{\bf k_2},{\bf k_3},{\bf k_4}) \equiv 
T^{\eta\to 0}_{\pi ,cc}({\bf k_1},{\bf k_2},{\bf k_3},{\bf k_4}) \nonumber\\
&&\qquad = 
\frac{\tilde \gamma H^8}{4 M^4}\Re\biggl\{
i \Bigl(
({\bf k_1} \cdot {\bf k_2})({\bf k_3} \cdot {\bf k_4}) +(\mbox{23 terms})
\Bigr)
\frac{\pi^2}{2^4}
\left( q_1 q_2 q_3 q_4 \right)^{-3/4} 2^3 
(\Gamma(1/4))^{-4} \nonumber \\
&&\qquad\qquad\qquad\qquad\qquad
\left(\frac{\pi}{8}\right)^2\int^0_{-\infty}d\eta' 
\left((-\eta')^{3/2}H_{3/4}^{(1)}(q_1 {\eta'}^2) \right)
\left((-\eta')^{3/2}H_{3/4}^{(1)}(q_2 {\eta'}^2) \right)
\nonumber\\ 
&&\qquad\qquad\qquad\qquad\qquad\qquad\qquad\qquad\qquad
\left((-\eta')^{3/2}H_{3/4}^{(1)}(q_3 {\eta'}^2) \right)
\left((-\eta')^{3/2}H_{3/4}^{(1)}(q_4 {\eta'}^2) \right)
\biggr\}
\nonumber\\
&&\qquad = 
\frac{\tilde \gamma H^8}{2^9 M^4} \left( \frac{\pi}{\Gamma(1/4)}\right)^4
\left( q_1 q_2 q_3 q_4 \right)^{-3/4}
\Bigl(
({\bf k_1} \cdot {\bf k_2})({\bf k_3} \cdot {\bf k_4}) +(\mbox{23 terms})
\Bigr)
\nonumber\\
&&\qquad\qquad
\Re\biggl\{ i
\int^0_{-\infty}d\eta' 
\left((-\eta')^{3/2}H_{3/4}^{(1)}(q_1 {\eta'}^2) \right)
\left((-\eta')^{3/2}H_{3/4}^{(1)}(q_2 {\eta'}^2) \right)
\nonumber\\ 
&&\qquad\qquad\qquad\qquad\qquad\qquad\qquad\qquad\qquad
\left((-\eta')^{3/2}H_{3/4}^{(1)}(q_3 {\eta'}^2) \right)
\left((-\eta')^{3/2}H_{3/4}^{(1)}(q_4 {\eta'}^2) \right)
\biggr\} ,
\label{eq:triccpi}
\end{eqnarray}
where $q_i$ is defined in eq.(\ref{qi})
and we have used
\begin{eqnarray}
&&\eta w_{-k_i}(\eta)=-\sqrt{\frac{\pi}{8}}(-\eta)^{3/2}
H_{3/4}^{(1)}(q_i \eta^2), \\
&&\eta w_{-k_i}^*|_{\eta\to 0} =-i\frac{\sqrt{\pi}}{2}
\left(\frac{q_i}{2}\right)^{-3/4} \frac{1}{\Gamma(1/4)}. 
\end{eqnarray}

Transforming variable as
\begin{eqnarray}
-\sqrt{q}\eta =x \qquad (q\equiv q_1=q_2=q_3=q_4) , 
\end{eqnarray}
the contact term contribution to the equilateral trispectrum is
\begin{eqnarray}
T_{\pi ,cc}(k,C_2,C_3,C_4)&\eq& 
\frac{\tilde \gamma H^8}{2^9 M^4}\left(\frac{\pi}{\Gamma(1/4)}\right)^4 q^{-3} 
\left(8 \left(\sum_{i=2,3,4}C_i^2\right) k^4\right)
\Re\left\{
i \int^0_{\infty} (-\frac{dx}{\sqrt{q}})\left(
\left(\frac{x}{\sqrt{q}} \right)^{3/2} H_{3/4}^{(1)}(x^2)
\right)^4
\right\}
\nonumber\\
&\eq& 
\sqrt{2}\frac{\tilde \gamma}{\alpha^{13/4}} H^4 
\left(\frac{H}{M}\right)^{-\frac{5}{2}}
\left(\frac{\pi}{\Gamma(1/4)}\right)^4 
\left( \sum_{i=2,3,4}C_i^2\right)   k^{-9}
\Re\left\{
i \int_0^{\infty} dx\left(
x^{3/2} H_{3/4}^{(1)}(x^2)
\right)^4
\right\} ,
\end{eqnarray}
where $C_i$ is defined in eq.(\ref{eq:C_i}) and $C_i$s are 
constrained by the momentum conservation as shown in (\ref{momcon}).

By using the relation~(\ref{zeta-pi-relation}) 
between $\zeta$ and $\pi$, we obtain the contact term contribution to
the trispectrum of curvature perturbation as 
\begin{eqnarray}
T_{\zeta , cc}(k,C_2,C_3,C_4)&\eq&\left( \frac{H}{M} \right)^4 
\frac{T_{\pi ,cc}(k,C_2,C_3,C_4)}{M^4}
\nonumber\\
&\eq& 
\sqrt{2} \frac{\tilde \gamma}{\alpha^{13/4}} \left(\frac{H}{M}\right)^{11/2}
\left(\frac{\pi}{\Gamma(1/4)}\right)^4 \left( \sum_{i=2,3,4} C_i^2 \right) k^{-9}
\Re\left\{
i \int_0^{\infty} dx\left(
x^{3/2} H_{3/4}^{(1)}(x^2)
\right)^4
\right\}.
\nonumber\\
\end{eqnarray}

In order to tame the asymptotic behavior of the integrand at
$x\to\infty$, we alter the integral route. The asymptotic expansion of
Hunkel function is 
\begin{eqnarray}
H_\nu^{(1)}(z) \to \sqrt{\frac{2}{\pi z}}exp\left[ 
i\left( z-\frac{2\nu +1}{4}\pi
\right)
\right]
\qquad(z\to \infty).
\end{eqnarray}
Thus, displacing the integral route from the real axis to an angle of
$\pi/4$ does not change the value of integral. This change of the
integral route is consistent with the $i\epsilon$ prescription in the
in-in formalism and picks up the correct vacuum. (See
(\ref{eqn:iepsilon}).) 
We define this axis as y-axis, which means
\begin{eqnarray}
x=\exp\left( \frac{\pi}{4} i  \right) y.
\end{eqnarray}
Under this transformation of variable, we have
\begin{eqnarray}
x^{3/2}H_{3/4}^{(1)}(x^2)
=\frac{2}{i\pi}y^{3/2} K_{3/4}(y^2).
\label{eq:H(1)3/4}
\end{eqnarray}
The expression (\ref{H/M}) for the power spectrum ${\cal P}_{\zeta}(k)$
can be inverted to express $H/M$ in terms of the power spectrum as 
\begin{eqnarray}
\frac{H}{M} = \left(
\pi (\Gamma(1/4))^2 \alpha ^{3/4} {\cal P}_{\zeta}(k)
\right)^{2/5}.
\label{calP}
\end{eqnarray}
Hence, we obtain
\begin{eqnarray}
T_{\zeta , cc}(k,C_2,C_3,C_4)
&\eq& 
\left(16 \sqrt{2} \pi^{11/5} \bigl(\Gamma(1/4) \bigr)^{2/5} 
\Re\left\{
i \int_0^{\infty}dy \exp\left( \frac{\pi}{4} i  \right)\left(
y^{3/2} K_{3/4}(y^2)
\right)^4
\right\} 
\right) \nonumber\\
&& \qquad\qquad\qquad\qquad\times
\bigl( {\cal P}_\zeta(k)\bigr)^{11/5}
\frac{\tilde \gamma}{\alpha ^{8/5}} \left( \sum_{i=2,3,4} C_i^2 \right) k^{-9}
\nonumber\\
&\eq&
-2.215\times 10^{-17} \times\left(
\frac{\bigl( {\cal P}_\zeta(k)\bigr)^{1/2}}{4.8 \times 10^-5}\right)^{22/5}
\frac{\tilde \gamma}{\alpha ^{8/5}} \left( \sum_{i=2,3,4} C_i^2 \right) k^{-9}.
\label{eq:Tzeta}
\end{eqnarray}

In the equilateral case where eq.~(\ref{eq:k-equal}) is satisfied, 
$\tau_{NL}^{cc}$ becomes
\begin{eqnarray}
\tau_{NL}^{cc}(k,C_2,C_3,C_4)&\eq& \frac{k^9}{\sqrt{2}}
\left[ \sum _{i=2,3,4} (1+C_i)^{-3/2} \right]^{-1}
 \frac{T_{\zeta, cc}(k)}{(2 \pi^2 {\cal P}_{\zeta}(k))^3}
\nonumber\\
&\eq&
\frac{k^9}{\sqrt{2}}
\left[ \sum _{i=2,3,4} (1+C_i)^{-3/2} \right]^{-1} (2 \pi^2 {\cal P}_{\zeta}(k))^{-3}
\sqrt{2} \frac{\tilde \gamma}{\alpha^{13/4}} \left(\frac{H}{M}\right)^{11/2}
\left(\frac{\pi}{\Gamma(1/4)}\right)^4 \left( \sum_{i=2,3,4} C_i^2 \right) k^{-9}
\nonumber\\
&&\qquad\qquad\qquad\qquad\qquad\qquad\qquad\qquad\qquad\qquad\qquad\qquad\times
\Re\left\{
i \int_0^{\infty} dx\left(
x^{3/2} H_{3/4}^{(1)}(x^2)
\right)^4
\right\}
\nonumber\\
&\eq&
\frac{1}{8 \pi^2(\Gamma(1/4))^4} \frac{\tilde \gamma}{\alpha^{13/4}}
\left(\frac{H}{M}\right)^{11/2} 
 ({\cal P}_{\zeta}(k))^{-3}
\Re\left\{
i \int_0^{\infty} dx\left(
x^{3/2} H_{3/4}^{(1)}(x^2)
\right)^4
\right\}\nonumber \\
&&\qquad\qquad\qquad\qquad\qquad\qquad\qquad\qquad\qquad\qquad\qquad\times
\frac{ \sum_{i=2,3,4} C_i^2}{ \sum_{i=2,3,4}\left(1+ C_i \right)^{-3/2}}.
\label{gnl}
\end{eqnarray}

Substituting eq.~(\ref{calP}) into eq.~(\ref{gnl}), we obtain
\begin{eqnarray}
\tau_{NL}^{cc}(k,C_2,C_3,C_4)&\eq&
\frac{1}{8 \pi^2(\Gamma(1/4))^4} \frac{\tilde \gamma}{\alpha^{13/4}}
\left(\pi (\Gamma(1/4))^2 \alpha ^{3/4} {\cal P}_{\zeta}(k)\right)^{11/5}
 ({\cal P}_{\zeta}(k))^{-3}
\Re\left\{
i \int_0^{\infty} dx\left(
x^{3/2} H_{3/4}^{(1)}(x^2)
\right)^4
\right\}\nonumber\\
&&\qquad\qquad\qquad\qquad\qquad\qquad\qquad\qquad\qquad\qquad\qquad\qquad\qquad\times
\frac{ \sum_{i=2,3,4} C_i^2}{ \sum_{i=2,3,4}\left(1+ C_i \right)^{-3/2}}.
\nonumber \\ 
&\eq& 
\frac{\tilde \gamma}{\alpha^{8/5}}({\cal P}_{\zeta}(k))^{-4/5}
\left(
\frac{1}{8} \pi^{1/5}(\Gamma(1/4))^{2/5}
 \Re\left\{
i \int_0^{\infty} dx\left(
x^{3/2} H_{3/4}^{(1)}(x^2)
\right)^4
\right\}
\right) \frac{ \sum_{i=2,3,4} C_i^2}{ \sum_{i=2,3,4}\left(1+ C_i \right)^{-3/2}}.
\nonumber\\
&\eq& 
\frac{\tilde \gamma}{\alpha^{8/5}}({\cal P}_{\zeta}(k))^{-4/5}
\left(
2 \pi^{-19/5}(\Gamma(1/4))^{2/5}
\Re\left\{
i \int_0^{\infty}dy \exp\left( \frac{\pi}{4} i  \right)\left(
y^{3/2} K_{3/4}(y^2)
\right)^4
\right\}
\right)\nonumber\\
&&\qquad\qquad\qquad\qquad\qquad\qquad\qquad\qquad\qquad\qquad\qquad\qquad\qquad\times
 \frac{ \sum_{i=2,3,4} C_i^2}{ \sum_{i=2,3,4}\left(1+ C_i \right)^{-3/2}}
\nonumber \\
&\!\! \simeq  \!\!&
-1.665\times10^{5} \times  \frac{\tilde \gamma}{\alpha^{8/5}}
\left(\frac{({\cal P}_{\zeta}(k))^{1/2}}{4.8\times10^{-5}}\right)^{-8/5}
\frac{ \sum_{i=2,3,4} C_i^2}{ \sum_{i=2,3,4}\left(1+ C_i \right)^{-3/2}} .
\label{taucc}
\end{eqnarray}

\subsection{scalar exchange contribution}

The scalar exchange contribution to the four point function is defined
as 
\begin{eqnarray}
&&\!\!\!\!\left\langle \pi_{\bf k_1}\pi_{\bf k_2}
\pi_{\bf k_3}\pi_{\bf k_4}\right\rangle_{se}
\nonumber\\
&&\equiv - \int^t_{-\infty} dt_2 \int^{t_2}_{-\infty} dt_1
\bigg\langle \biggl[ \int dx_1^3 \frac{\beta a(t_1)}{2M^2} 
\left( \frac{d}{d t_1}\pi(t_1,x_1) \right)\left( \nabla \pi(t_1,x_1)\right)^2, 
\nonumber \\
&&\qquad\qquad\qquad\qquad\qquad\qquad
\Bigl[\int dx_2^3 \frac{\beta a(t_2)}{2M^2} 
\left( \frac{d}{d t_2}\pi(t_2,x_2) \right)\left( \nabla \pi(t_2,x_2)\right)^2, 
\pi_{\bf k_1}\pi_{\bf k_2}\pi_{\bf k_3}\pi_{\bf k_4} \Bigr]\biggr] \bigg\rangle
\nonumber \\
&&= - \int^t_{-\infty} dt_2 \int^{t_2}_{-\infty} dt_1
\Biggl\{ \bigg\langle \left( \int dx_1^3 \frac{\beta a(t_1)}{2M^2} 
\left( \frac{d}{d t_1}\pi(t_1,x_1)\right)\left( \nabla \pi(t_1,x_1)\right)^2 \right) \nonumber \\
&&\qquad\qquad\qquad\qquad\qquad\qquad\qquad\qquad
\left( \int dx_2^3 \frac{\beta a(t_2)}{2M^2} 
\left( \frac{d}{d t_2}\pi(t_2,x_2)\right)\left( \nabla \pi(t_2,x_2)\right)^2 \right)
\pi_{\bf k_1}\pi_{\bf k_2}\pi_{\bf k_3}\pi_{\bf k_4}\bigg\rangle
\nonumber \\
&&\qquad\qquad \qquad\qquad\qquad
+\bigg\langle \pi_{\bf k_1}\pi_{\bf k_2}\pi_{\bf k_3}\pi_{\bf k_4}
\left( \int dx_2^3 \frac{\beta a(t_2)}{2M^2} 
\left( \frac{d}{d t_2}\pi(t_2,x_2)\right)\left( \nabla \pi(t_2,x_2)\right)^2 \right) 
\nonumber \\
&&\qquad\qquad\qquad\qquad\qquad\qquad\qquad\qquad\qquad\qquad
 \left( \int dx_1^3 \frac{\beta a(t_1)}{2M^2} 
\left( \frac{d}{d t_1}\pi(t_1,x_1)\right)
\left( \nabla \pi(t_1,x_1)\right)^2 \right) \bigg\rangle
\Biggr\}
\nonumber \\
&&\qquad
+\int^t_{-\infty} dt_2 \int^{t}_{-\infty} dt_1
\bigg\langle \left( \int dx_1^3 \frac{\beta a(t_1)}{2M^2} 
\left( \frac{d}{d t_1}\pi(t_1,x_1)\right)
\left( \nabla \pi(t_1,x_1)\right)^2 \right) \nonumber \\
&&\qquad\qquad\qquad\qquad\qquad\qquad\qquad
\pi_{\bf k_1}\pi_{\bf k_2}\pi_{\bf k_3}\pi_{\bf k_4}
\left( \int dx_2^3 \frac{\beta a(t_2)}{2M^2} 
\left( \frac{d}{d t_2}\pi(t_2,x_2)\right)
\left( \nabla \pi(t_2,x_2)\right)^2 \right)
\bigg\rangle .
\end{eqnarray}
Using eqs.~(\ref{eq:transform}), (\ref{eq:u_k}) and (\ref{eq:exchange}),
we can transform this as 
\begin{eqnarray}
&&\!\!\!\!\left\langle \pi_{\bf k_1}\pi_{\bf k_2}
\pi_{\bf k_3}\pi_{\bf k_4}\right\rangle_{se}
\nonumber\\
&& = 
\frac{\beta^2}{4M^4} \Biggl\{
-\int^\eta_{-\infty} a(\eta_2) d\eta_2 \int^{\eta_2}_{-\infty} a(\eta_1) d\eta_1
\Bigg\langle
\int dx_1^3\left( \frac{d}{d\eta_1}\frac{u(\eta_1,x_1)}{a(\eta_1)}\right) 
\left(\frac{\nabla u(\eta_1,x_1)}{a(\eta_1)}\right) ^2
\nonumber \\
&& \qquad\qquad\qquad \times
\int dx_2^3\left( \frac{d}{d\eta_2}\frac{u(\eta_2,x_2)}{a(\eta_2)}\right) 
\left(\frac{\nabla u(\eta_2,x_2)}{a(\eta_2)}\right) ^2
\frac{u_{\bf k_1}(\eta)}{a(\eta)}\frac{u_{\bf k_2}(\eta)}{a(\eta)}
\frac{u_{\bf k_3}(\eta)}{a(\eta)}\frac{u_{\bf k_4}(\eta)}{a(\eta)}
+\mbox{h.c.}
\Bigg\rangle
\nonumber \\
&&\qquad\qquad
+\int^\eta_{-\infty} a(\eta_2) d\eta_2 \int^{\eta}_{-\infty} a(\eta_1) d\eta_1
\Bigg\langle
\int dx_1^3\left(\frac{d}{d\eta_1}\frac{u(\eta_1,x_1)}{a(\eta_1)}\right) 
\left(\frac{\nabla u(\eta_1,x_1)}{a(\eta_1)}\right) ^2
\nonumber \\
&& \qquad\qquad\qquad \times
\frac{u_{\bf k_1}(\eta)}{a(\eta)}\frac{u_{\bf k_2}(\eta)}{a(\eta)}
\frac{u_{\bf k_3}(\eta)}{a(\eta)}\frac{u_{\bf k_4}(\eta)}{a(\eta)}
\int dx_2^3\left(\frac{d}{d\eta_2}\frac{u(\eta_2,x_2)}{a(\eta_2)}\right) 
\left(\frac{\nabla u(\eta_2,x_2)}{a(\eta_2)}\right) ^2
\Bigg\rangle
\Biggr\}
\nonumber
\end{eqnarray}
\begin{eqnarray}
&&= 
\frac{\beta^2 H^8}{4M^4}(2\pi)^3 
\delta^3({\bf k_1}+{\bf k_2}+{\bf k_3}+{\bf k_4}) \times 
\nonumber \\ 
&& \quad \biggl\{ 
-\int^{\eta}_{-\infty} \frac{d\eta_2}{\eta_2}\int^{\eta_2}_{-\infty}
\frac{d\eta_1}{\eta_1} \biggl( \bigl\{
4({\bf k_1} \cdot {\bf k_2})({\bf k_3} \cdot {\bf k_4})
\bigl(-\eta_1w_{-k_1}(\eta_1)\bigr)\bigl(-\eta_1w_{-k_2}(\eta_1)\bigr)
\left(\frac{d}{d\eta_1} 
\bigl(-\eta_1w_{\left|{\bf k_1}+{\bf k_2}\right|}(\eta_1)\bigr) \right)
\nonumber \\
&& \qquad\qquad\qquad\qquad\qquad\quad \times
\bigl(-\eta_2w_{-k_3}(\eta_2)\bigr)
\bigl(-\eta_2w_{-k_4}(\eta_2)\bigr)
\left(\frac{d}{d\eta_2} 
\bigl(-\eta_2w_{\left|{\bf k_1}+{\bf k_2}\right|}^*(\eta_2)\bigr)\right)
\nonumber \\
&& \qquad\qquad\qquad\qquad\qquad\quad \times
\bigl(-\eta w_{-k_1}^*(\eta)\bigr)\bigl(-\eta w_{-k_2}^*(\eta)\bigr)
\bigl(-\eta w_{-k_3}^*(\eta)\bigr)
\bigl(-\eta w_{-k_4}^*(\eta)\bigr)
\qquad +5 \mbox{terms} \bigr\}
\nonumber \\ 
&& \qquad\qquad\quad
+\bigl\{
4(-k_1^2 -{\bf k_1} \cdot {\bf k_2})({\bf k_3} \cdot {\bf k_4})
\bigl(-\eta_1w_{-k_1}(\eta_1)\bigr)
\left(\frac{d}{d\eta_1} \bigl(-\eta_1w_{-k_2}(\eta_1)\bigr)\right)
\bigl(-\eta_1w_{\left|{\bf k_1}+{\bf k_2}\right|}(\eta_1)\bigr)
\nonumber \\
&& \qquad\qquad\qquad\qquad\qquad\quad \times
\bigl(-\eta_2w_{-k_3}(\eta_2)\bigr)
\bigl(-\eta_2w_{-k_4}(\eta_2)\bigr)
\left(\frac{d}{d\eta_2} 
\bigl(-\eta_2w_{\left|{\bf k_1}+{\bf k_2}\right|}^*(\eta_2)\bigr) \right)
\nonumber \\
&& \qquad\qquad\qquad\qquad\qquad\quad \times
\bigl(-\eta w_{-k_1}^*(\eta)\bigr)\bigl(-\eta w_{-k_2}^*(\eta)\bigr)
\bigl(-\eta w_{-k_3}^*(\eta)\bigr)
\bigl(-\eta w_{-k_4}^*(\eta)\bigr)
\qquad
+11 \mbox{terms} \bigr\}
\nonumber \\
&& \qquad\qquad\quad
+\bigl\{
4({\bf k_1} \cdot {\bf k_2})(-k_3^2-{\bf k_3} \cdot {\bf k_4})
\bigl(-\eta_1w_{-k_1}(\eta_1)\bigr)\bigl(-\eta_1w_{-k_2}(\eta_1)\bigr)
\left(\frac{d}{d\eta_1} 
\bigl(-\eta_1w_{\left|{\bf k_1}+{\bf k_2}\right|}(\eta_1)\bigr)\right)
\nonumber \\
&& \qquad\qquad\qquad\qquad\qquad\quad \times
\bigl(-\eta_2w_{-k_3}(\eta_2)\bigr)
\left(\frac{d}{d\eta_2} \bigl(-\eta_2w_{-k_4}(\eta_2)\bigr)\right)
\bigl(-\eta_2w_{\left|{\bf k_1}+{\bf k_2}\right|}^*(\eta_2)\bigr) 
\nonumber \\
&& \qquad\qquad\qquad\qquad\qquad\quad \times
\bigl(-\eta w_{-k_1}^*(\eta)\bigr)\bigl(-\eta w_{-k_2}^*(\eta)\bigr)
\bigl(-\eta w_{-k_3}^*(\eta)\bigr)
\bigl(-\eta w_{-k_4}^*(\eta)\bigr)
\qquad
+ 11 \mbox{terms} \bigr\}
\nonumber \\
&& \qquad\qquad\quad
+\bigl\{
4(-k_1^2 -{\bf k_1} \cdot {\bf k_2})(-k_3^2-{\bf k_3} \cdot {\bf k_4})
\bigl(-\eta_1w_{-k_1}(\eta_1)\bigr)
\left(\frac{d}{d\eta_1} \bigl(-\eta_1w_{-k_2}(\eta_1)\bigr)\right)
\bigl(-\eta_1w_{\left|{\bf k_1}+{\bf k_2}\right|}(\eta_1)\bigr)
\nonumber \\
&& \qquad\qquad\qquad\qquad\qquad\quad \times
\bigl(-\eta_2w_{-k_3}(\eta_2)\bigr)
\left(\frac{d}{d\eta_2} \bigl(-\eta_2w_{-k_4}(\eta_2)\bigr)\right)
\bigl(-\eta_2w_{\left|{\bf k_1}+{\bf k_2}\right|}^*(\eta_2)\bigr) 
\nonumber \\
&& \qquad\qquad\qquad\qquad\qquad\quad \times
\bigl(-\eta w_{-k_1}^*(\eta)\bigr)\bigl(-\eta w_{-k_2}^*(\eta)\bigr)
\bigl(-\eta w_{-k_3}^*(\eta)\bigr)
\bigl(-\eta w_{-k_4}^*(\eta)\bigr)
\qquad
+ 23 \mbox{terms}\bigr\} +\mbox{c.c.} 
\biggr)
\nonumber \\
&&\qquad
+\int^{\eta}_{-\infty} \frac{d\eta_1}{\eta_1}\int^{\eta}_{-\infty}
\frac{d\eta_2}{\eta_2} \biggl( \Bigl[ \bigl\{
4({\bf k_1} \cdot {\bf k_2})({\bf k_3} \cdot {\bf k_4})
\bigl(-\eta_1w_{-k_1}(\eta_1)\bigr)\bigl(-\eta_1w_{-k_2}(\eta_1)\bigr)
\left(\frac{d}{d\eta_1} 
\bigl(-\eta_1w_{\left|{\bf k_1}+{\bf k_2}\right|}(\eta_1)\bigr)\right)
\nonumber \\
&& \qquad\qquad\qquad\qquad\qquad\quad \times
\bigl(-\eta w_{-k_1}^*(\eta)\bigr)
\bigl(-\eta w_{-k_2}^*(\eta)\bigr)\bigl(-\eta w_{k_3}(\eta)\bigr)
\bigl(-\eta w_{k_4}(\eta)\bigr)
\nonumber \\
&& \qquad\qquad\qquad\qquad\qquad\quad \times
\bigl(-\eta_2w_{k_3}^*(\eta_2)\bigr)
\bigl(-\eta_2w_{k_4}^*(\eta_2)\bigr)
\left(\frac{d}{d\eta_2} 
\bigl(-\eta_2w_{\left|{\bf k_1}+{\bf k_2}\right|}^*(\eta_2)\bigr) \right)
\qquad
+ 5 \mbox{terms} \bigr\}
\nonumber \\ 
&& \qquad\qquad\quad
+\bigl\{
4(-k_1^2 -{\bf k_1} \cdot {\bf k_2})({\bf k_3} \cdot {\bf k_4})
\bigl(-\eta_1w_{-k_1}(\eta_1)\bigr)
\bigl(-\eta_1w_{\left|{\bf k_1}+{\bf k_2}\right|}(\eta_1)\bigr)
\left(\frac{d}{d\eta_1} \bigl(-\eta_1w_{-k_2}(\eta_1)\bigr)\right)
\nonumber \\
&& \qquad\qquad\qquad\qquad\qquad\quad \times
\bigl(-\eta w_{-k_1}^*(\eta)\bigr)
\bigl(-\eta w_{-k_2}^*(\eta)\bigr)\bigl(-\eta w_{k_3}(\eta)\bigr)
\bigl(-\eta w_{k_4}(\eta)\bigr)
\nonumber \\
&& \qquad\qquad\qquad\qquad\qquad\quad \times
\bigl(-\eta_2w_{k_3}^*(\eta_2)\bigr)
\bigl(-\eta_2w_{k_4}^*(\eta_2)\bigr)
\left(\frac{d}{d\eta_2} 
\bigl(-\eta_2w_{\left|{\bf k_1}+{\bf k_2}\right|}^*(\eta_2)\bigr) \right)
\qquad
+ 11 \mbox{terms} \bigr\}
\nonumber \\
&& \qquad\qquad\quad
+\bigl\{
4({\bf k_1} \cdot {\bf k_2})(-k_3^2-{\bf k_3} \cdot {\bf k_4})
\bigl(-\eta_1w_{-k_1}(\eta_1)\bigr)\bigl(-\eta_1w_{-k_2}(\eta_1)\bigr)
\left(\frac{d}{d\eta_1} 
\bigl(-\eta_1w_{\left|{\bf k_1}+{\bf k_2}\right|}(\eta_1)\bigr)\right)
\nonumber \\
&& \qquad\qquad\qquad\qquad\qquad\quad \times
\bigl(-\eta w_{-k_1}^*(\eta)\bigr)
\bigl(-\eta w_{-k_2}^*(\eta)\bigr)\bigl(-\eta w_{k_3}(\eta)\bigr)
\bigl(-\eta w_{k_4}(\eta)\bigr)
\nonumber \\
&& \qquad\qquad\qquad\qquad\qquad\quad \times
\bigl(-\eta_2w_{k_3}^*(\eta_2)\bigr)
\bigl(-\eta_2w_{\left|{\bf k_1}+{\bf k_2}\right|}^*(\eta_2)\bigr) 
\left(\frac{d}{d\eta_2} \bigl(-\eta_2w_{k_4}^*(\eta_2)\bigr) \right)
\qquad
+ 11 \mbox{terms} \bigr\}
\nonumber \\
&& \qquad\qquad\quad
+\bigl\{
4(-k_1^2 -{\bf k_1} \cdot {\bf k_2})(-k_3^2-{\bf k_3} \cdot {\bf k_4})
\bigl(-\eta_1w_{-k_1}(\eta_1)\bigr)
\bigl(-\eta_1w_{\left|{\bf k_1}+{\bf k_2}\right|}(\eta_1)\bigr)
\left(\frac{d}{d\eta_1} \bigl(-\eta_1w_{-k_2}(\eta_1)\bigr)\right)
\nonumber \\
&& \qquad\qquad\qquad\qquad\qquad\quad \times
\bigl(-\eta w_{-k_1}^*(\eta)\bigr)
\bigl(-\eta w_{-k_2}^*(\eta)\bigr)\bigl(-\eta w_{k_3}(\eta)\bigr)
\bigl(-\eta w_{k_4}(\eta)\bigr)
\nonumber \\
&& \qquad\qquad\qquad\qquad\qquad\quad \times
\bigl(-\eta_2w_{k_3}^*(\eta_2)\bigr)
\bigl(-\eta_2w_{\left|{\bf k_1}+{\bf k_2}\right|}^*(\eta_2)\bigr) 
\left(\frac{d}{d\eta_2} \bigl(-\eta_2w_{k_4}^*(\eta_2)\bigr) \right)
\qquad
+ 23 \mbox{terms} \bigr\} \Bigr] \biggr)\biggr\}
\nonumber
\end{eqnarray}
\begin{eqnarray}
&&= 
\frac{\beta^2 H^8}{4M^4}(2\pi)^3 
\delta^3({\bf k_1}+{\bf k_2}+{\bf k_3}+{\bf k_4}) \times 
\nonumber \\ 
&& \quad \biggl\{  -
\int^{\eta}_{-\infty} \frac{d\eta_2}{\eta_2}\int^{\eta_2}_{-\infty}
\frac{d\eta_1}{\eta_1}\Bigl(   \bigl\{ \bigl[
({\bf k_1} \cdot {\bf k_2})
\bigl(-\eta_1w_{-k_1}(\eta_1)\bigr)\bigl(-\eta_1w_{-k_2}(\eta_1)\bigr)
\left(\frac{d}{d\eta_1}
\bigl(-\eta_1w_{\left|{\bf k_1}+{\bf k_2}\right|}(\eta_1)\bigr)\right)
\nonumber \\
&& \qquad\qquad\quad\qquad\qquad\qquad\qquad
-
2(k_1^2 +{\bf k_1} \cdot {\bf k_2})
\bigl(-\eta_1w_{-k_1}(\eta_1)\bigr)
\bigl(-\eta_1w_{\left|{\bf k_1}+{\bf k_2}\right|}(\eta_1)\bigr)
\left(\frac{d}{d\eta_1}\bigl(-\eta_1w_{-k_2}(\eta_1)\bigr)\right)\bigr]
\nonumber \\
&& \qquad\qquad\qquad\qquad\qquad\times \bigl[
({\bf k_3} \cdot {\bf k_4})\bigl(-\eta_2w_{-k_3}(\eta_2)\bigr)
\bigl(-\eta_2w_{-k_4}(\eta_2)\bigr)
\left(\frac{d}{d\eta_2}
\bigl(-\eta_2w_{\left|{\bf k_1}+{\bf k_2}\right|}^*(\eta_2)\bigr)\right) 
\nonumber \\
&& \qquad\qquad\quad\qquad\qquad\qquad\qquad
-2(k_3^2+{\bf k_3} \cdot {\bf k_4})\bigl(-\eta_2w_{-k_3}(\eta_2)\bigr)
\left(\frac{d}{d\eta_2}\bigl(-\eta_2w_{-k_4}(\eta_2)\bigr)\right)
\bigl(-\eta_2w_{\left|{\bf k_1}+{\bf k_2}\right|}^*(\eta_2)\bigr)\bigr]
\nonumber \\
&& \qquad\qquad\qquad\qquad\qquad \times
\bigl(-\eta w_{-k_1}^*(\eta)\bigr)\bigl(-\eta w_{-k_2}^*(\eta)\bigr)
\bigl(-\eta w_{-k_3}^*(\eta)\bigr)
\bigl(-\eta w_{-k_4}^*(\eta)\bigr) 
\nonumber \\
&&\qquad\qquad\qquad\qquad\qquad\qquad\qquad\qquad\qquad\qquad\qquad
\qquad\qquad\qquad
+\mbox{23 permutational terms}\bigr\}+ c.c.\Bigr) 
\nonumber \\
&&\qquad
+\biggl[\biggl(\int^{\eta}_{-\infty} \frac{d\eta_1}{\eta_1} \bigl\{
({\bf k_1} \cdot {\bf k_2})
\bigl(-\eta_1w_{-k_1}(\eta_1)\bigr)\bigl(-\eta_1w_{-k_2}(\eta_1)\bigr)
\left(\frac{d}{d\eta_1}
\bigl(-\eta_1w_{\left|{\bf k_1}+{\bf k_2}\right|}(\eta_1)\bigr)\right)
\nonumber \\
&& \qquad\qquad\quad\qquad\qquad\qquad\qquad
-
2(k_1^2 +{\bf k_1} \cdot {\bf k_2})
\bigl(-\eta_1w_{-k_1}(\eta_1)\bigr)
\bigl(-\eta_1w_{\left|{\bf k_1}+{\bf k_2}\right|}(\eta_1)\bigr)
\left(\frac{d}{d\eta_1}\bigl(-\eta_1w_{-k_2}(\eta_1)\bigr)\right)
\bigr\}\biggr)
\nonumber\\
&&\qquad\quad \times
\biggl(\int^{\eta}_{-\infty} \frac{d\eta_2}{\eta_2} \bigl\{
({\bf k_3} \cdot {\bf k_4})\bigl(-\eta_2w_{-k_3}^*(\eta_2)\bigr)
\bigl(-\eta_2w_{-k_4}^*(\eta_2)\bigr)
\left(\frac{d}{d\eta_2}
\bigl(-\eta_2w_{\left|{\bf k_1}+{\bf k_2}\right|}^*(\eta_2)\bigr)\right)
\nonumber \\
&& \qquad\qquad\quad\qquad\qquad\qquad\qquad
-2(k_3^2+{\bf k_3} \cdot {\bf k_4})\bigl(-\eta_2w_{-k_3}^*(\eta_2)\bigr)
\left(\frac{d}{d\eta_2} \bigl(-\eta_2w_{-k_4}^*(\eta_2)\bigr)\right)
\bigl(-\eta_2w_{\left|{\bf k_1}+{\bf k_2}\right|}^*(\eta_2)\bigr) 
\bigr\}\biggr)
\nonumber \\
&& \qquad\qquad\qquad\qquad\qquad \times
\bigl(-\eta w_{-k_1}^*(\eta)\bigr)\bigl(-\eta w_{-k_2}^*(\eta)\bigr)
\bigl(-\eta w_{k_3}(\eta)\bigr)
\bigl(-\eta w_{k_4}(\eta)\bigr) 
\nonumber \\
&&\qquad\qquad\qquad\qquad\qquad\qquad\qquad\qquad\qquad\qquad\qquad
\qquad\qquad\qquad\qquad
+\mbox{23 permutational terms} \biggr]\biggr\}
\nonumber\\
&&= 
\frac{\beta^2 \alpha H^{10} }{4M^6}\left(\frac{\pi}{8}\right)^3(2\pi)^3 
\delta^3({\bf k_1}+{\bf k_2}+{\bf k_3}+{\bf k_4}) 
 \biggl\{  -
\int^{\eta}_{-\infty}d\eta_2\int^{\eta_2}_{-\infty}d\eta_1
\nonumber \\ 
&& \quad
\Bigl(   \bigl\{ \bigl[
({\bf k_1} \cdot {\bf k_2}) (k_1^2 +k_2^2 +2 {\bf k_1} \cdot {\bf k_2})
\bigl( (-\eta_1)^{3/2} H_{3/4}^{(1)} (q_1 \eta_1^2) \bigr)
\bigl( (-\eta_1)^{3/2} H_{3/4}^{(1)} (q_2 \eta_1^2) \bigr)
\bigl( (-\eta_1)^{3/2} H_{-1/4}^{(1)} (q_{12} \eta_1^2) \bigr)
\nonumber \\
&& \qquad\qquad\quad
-
2(k_1^2 +{\bf k_1} \cdot {\bf k_2}) k_2^2
\bigl( (-\eta_1)^{3/2} H_{3/4}^{(1)} (q_1 \eta_1^2) \bigr)
\bigl( (-\eta_1)^{3/2} H_{-1/4}^{(1)} (q_2 \eta_1^2) \bigr)
\bigl( (-\eta_1)^{3/2} H_{3/4}^{(1)} (q_{12} \eta_1^2) \bigr)
\bigr]
\nonumber \\
&&\qquad\times 
\bigl[
({\bf k_3} \cdot {\bf k_4})(k_3^2 +k_4^2 +2 {\bf k_3} \cdot {\bf k_4})
\bigl( (-\eta_2)^{3/2} H_{3/4}^{(1)} (q_3 \eta_2^2) \bigr)
\bigl( (-\eta_2)^{3/2} H_{3/4}^{(1)} (q_4 \eta_2^2) \bigr)
\bigl( (-\eta_2)^{3/2} H_{-1/4}^{(2)} (q_{34} \eta_2^2) \bigr) 
\nonumber \\
&& \qquad\qquad\quad
-
2(k_3^2 +{\bf k_3} \cdot {\bf k_4}) k_4^2
\bigl( (-\eta_2)^{3/2} H_{3/4}^{(1)} (q_3 \eta_2^2) \bigr)
\bigl( (-\eta_2)^{3/2} H_{-1/4}^{(1)} (q_4 \eta_2^2) \bigr)
\bigl( (-\eta_2)^{3/2} H_{3/4}^{(2)} (q_{34} \eta_2^2) \bigr)
\bigr]
\nonumber \\
&& \qquad\qquad \times
\bigl(-\eta w_{-k_1}^*(\eta)\bigr)\bigl(-\eta w_{-k_2}^*(\eta)\bigr)
\bigl(-\eta w_{-k_3}^*(\eta)\bigr)
\bigl(-\eta w_{-k_4}^*(\eta)\bigr) 
\nonumber \\
&&\qquad\qquad\qquad\qquad\qquad\qquad\qquad\qquad
\qquad\qquad\qquad
+\mbox{23 permutational terms}\bigr\}+ c.c.\Bigr) 
\nonumber \\
&&
+\biggl[\biggl(\int^{\eta}_{-\infty} d\eta_1 \bigl[
({\bf k_1} \cdot {\bf k_2}) (k_1^2 +k_2^2 +2 {\bf k_1} \cdot {\bf k_2})
\bigl( (-\eta_1)^{3/2} H_{3/4}^{(1)} (q_1 \eta_1^2) \bigr)
\bigl( (-\eta_1)^{3/2} H_{3/4}^{(1)} (q_2 \eta_1^2) \bigr)
\bigl( (-\eta_1)^{3/2} H_{-1/4}^{(1)} (q_{12} \eta_1^2) \bigr)
\nonumber \\
&& \quad\qquad\qquad\qquad
-
2(k_1^2 +{\bf k_1} \cdot {\bf k_2}) k_2^2
\bigl( (-\eta_1)^{3/2} H_{3/4}^{(1)} (q_1 \eta_1^2) \bigr)
\bigl( (-\eta_1)^{3/2} H_{-1/4}^{(1)} (q_2 \eta_1^2) \bigr)
\bigl( (-\eta_1)^{3/2} H_{3/4}^{(1)} (q_{12} \eta_1^2) \bigr)
\bigr]
\biggr)
\nonumber\\
&& \times
\biggl(\int^{\eta}_{-\infty} d\eta_2 \bigl[
({\bf k_3} \cdot {\bf k_4})(k_3^2 +k_4^2 +2 {\bf k_3} \cdot {\bf k_4})
\bigl( (-\eta_2)^{3/2} H_{3/4}^{(2)} (q_3 \eta_2^2) \bigr)
\bigl( (-\eta_2)^{3/2} H_{3/4}^{(2)} (q_4 \eta_2^2) \bigr)
\bigl( (-\eta_2)^{3/2} H_{-1/4}^{(2)} (q_{34} \eta_2^2) \bigr) 
\nonumber \\
&& \quad\qquad\qquad\qquad
-
2(k_3^2 +{\bf k_3} \cdot {\bf k_4}) k_4^2
\bigl( (-\eta_2)^{3/2} H_{3/4}^{(2)} (q_3 \eta_2^2) \bigr)
\bigl( (-\eta_2)^{3/2} H_{-1/4}^{(2)} (q_4 \eta_2^2) \bigr)
\bigl( (-\eta_2)^{3/2} H_{3/4}^{(2)} (q_{34} \eta_2^2) \bigr)
\bigr]
\biggr)
\nonumber \\
&& \qquad\qquad\qquad\qquad\qquad\qquad\qquad\qquad\qquad\qquad \times
\bigl(-\eta w_{-k_1}^*(\eta)\bigr)\bigl(-\eta w_{-k_2}^*(\eta)\bigr)
\bigl(-\eta w_{k_3}(\eta)\bigr)
\bigl(-\eta w_{k_4}(\eta)\bigr) 
\nonumber \\
&&\qquad\qquad\qquad\qquad\qquad\qquad\qquad\qquad\qquad\qquad\qquad
\qquad\qquad\qquad\qquad\qquad
+\mbox{23 permutational terms} \biggr]\biggr\} ,\label{eq:Tpise}
\end{eqnarray}
where $q_i$ and $q_{ij}$ are defined in eq.~(\ref{qi}) and
eq.~(\ref{qij}), respectively, and we have used the following identity:  
\begin{eqnarray}
\frac{d}{d\eta}
\left( (-\eta)^{3/2} H_{3/4}^{(i)}(q \eta^2) \right)
=
2\beta \eta
\left( (-\eta)^{3/2} H_{-1/4}^{(i)}(q \eta^2) \right) 
\qquad (i=1,2).
\end{eqnarray}
 
Now let us take the limit $\eta \to 0$ and consider the equilateral case
where the amplitudes of all momenta are the same. Then, we have 
\begin{eqnarray}
&&T_{\pi, se}(k,C_2,C_3,C_4) 
\nonumber\\
&&\quad=
\frac{2\beta^2_3 \alpha H^{10}k^8}{M^6} 
\left( \frac{\sqrt{\pi}}{2}\left(\frac{2}{q}\right)^{3/4}
\Gamma\left(1/4\right)^{-1} \right)^4 
\left(\frac{\pi}{8}\right)^3 q^{-11/2}
\sum_{i=2,3,4}
\biggl\{4(1+C_i)^2 \biggl[ -\biggl( \int^{0}_{\infty}dx_2
\int^{x_2}_{\infty}dx_1
\nonumber \\
&&\qquad\qquad\qquad\qquad\qquad\times
 x_1^{9/2} \Bigl( C_i \bigl( H_{3/4}^{(1)}(x_1^2) \bigr)^2 
H_{-1/4}^{(1)}(2(1+C_i)x_1^2) 
- H_{3/4}^{(1)}(x_1^2) H_{-1/4}^{(1)}(x_1^2) H_{3/4}^{(1)}(2(1+C_i)x_1^2) \Bigr)
\nonumber\\
&&\qquad\qquad\qquad\qquad\qquad\times
 x_2^{9/2} \Bigl( C_i \bigl( H_{3/4}^{(1)}(x_2^2) \bigr)^2 
H_{-1/4}^{(2)}(2(1+C_i)x_2^2) 
- H_{3/4}^{(1)}(x_2^2) H_{-1/4}^{(1)}(x_2^2) H_{3/4}^{(2)}(2(1+C_i)x_2^2) \Bigr)
+ \mbox{c.c.} \biggr)
\nonumber\\
&&\qquad\qquad
+\biggl|\int^{0}_{\infty}dx_1
x_1^{9/2} \Bigl( C_i \bigl( H_{3/4}^{(1)}(x_1^2) \bigr)^2 
H_{-1/4}^{(1)}(2(1+C_i)x_1^2) 
- H_{3/4}^{(1)}(x_1^2) H_{-1/4}^{(1)}(x_1^2) H_{3/4}^{(1)}(2(1+C_i)x_1^2) \Bigr)
\biggr|^2
\biggr]
\biggr\}
\nonumber \\
&&\quad=
\frac{2^7 \sqrt{2}}{\pi}\left(\Gamma( 1/4) \right)^{-4}
\beta^2 \alpha^{-13/4} H^{4}
\left( \frac{H}{M}\right)^{-\frac{5}{2}} k^{-9}
\sum_{i=2,3,4} (A(C_i)+B(C_i)),
\end{eqnarray}
where $C_i$, $A(C)$ and $B(C)$ are defined in eq.~(\ref{eq:C_i}) 
eq.~(\ref{A_i}) and eq.~(\ref{B_i}), respectively, 
and we have used eq.~(\ref{eq:H(1)3/4}) and 
\begin{eqnarray}
&&
x^{-1/2} H^{(1)}_{-1/4}(x^2)=
 \frac{2}{\pi i} y^{-1/2} K_{-1/4}(y^2), 
\\
&&x^{-1/2} H^{(2)}_{-1/4}(x^2)=
\sqrt{2} y^{-1/2}\left( I_{-1/4}(y^2) - i I_{1/4}(y^2) \right) ,
\label{1/4} \\
&&
x^{3/2} H^{(2)}_{3/4}(x^2)=
\sqrt{2} y^{3/2}\left( -I_{3/4}(y^2) +i I_{-3/4}(y^2) \right) .
\label{eq:Hankel(2)3/4}
\end{eqnarray}
Equations~(\ref{eq:Hankel(2)3/4}) and (\ref{1/4}) are special cases of
the following relation:
\begin{eqnarray}
H^{(2)}_{\nu}(iz)&\eq& J_{\nu}(iz)-iN_{\nu}(iz) 
\nonumber\\
&\eq& J_{\nu}(iz)-\frac{i\cos\nu\pi}{\sin\nu\pi}J_{\nu}(iz)
+\frac{i}{\sin\nu\pi}J_{-\nu}(iz)
\nonumber \\
&\eq&
\frac{i}{\sin\nu\pi}\left(-e^{i\nu\pi}J_{\nu}(iz)+J_{-\nu}(iz)\right)
\nonumber\\
&\eq&
\frac{i}{\sin\nu\pi}\left(-e^{i\nu\pi}e^{\frac{i}{2}\nu\pi}
\left(\frac{z}{2}\right)^\nu \sum_{n=0}^{\infty}
\frac{(z/2)^{2n}}{n!\Gamma(\nu+n+1)}
+e^{-\frac{i}{2}\nu\pi}
\left(\frac{z}{2}\right)^{-\nu} \sum_{n=0}^{\infty}
\frac{(z/2)^{2n}}{n!\Gamma(-\nu+n+1)}\right)
\nonumber\\
&\eq&
\frac{ie^{-\frac{i}{2}\nu\pi}}{\sin\nu\pi}\left(
-e^{2\nu\pi i} I_{\nu}(z) +I_{-\nu}(z)
\right) .
\end{eqnarray}

The scalar exchange contribution to the four-point function of curvature
perturbation is 
\begin{eqnarray}
T_{\zeta,se}(k,C_2,C_3,C_4)&\eq&\left(\frac{H}{M}\right)^4 
\frac{T_{\pi,se}(k,C_2,C_3,C_4)}{M^4}
\nonumber\\
&\eq& 
\frac{2^7\sqrt{2}}{\pi}\left(\Gamma( 1/4) \right)^{-4}\beta^2 \alpha^{-13/4} 
\left(\frac{H}{M}\right)^{11/2} k^{-9} \sum_{i=2,3,4}
(A(C_i)+B(C_i)) 
\nonumber\\
&\eq& 
2^{15/2} \pi^{6/5} \left(\Gamma( 1/4) \right)^{2/5} 
\bigl({\cal P}_\zeta (k)\bigr)^{11/5} \frac{\beta^2}{\alpha^{8/5}}
k^{-9} \sum_{i=2,3,4}
(A(C_i)+B(C_i)) 
\nonumber\\
&\eq& 
1.190 \times 10^{-16} 
\left(\frac{\bigl({\cal P}_\zeta (k)\bigr)^{1/2}}{4.8\times 10^{-5}}\right)^{22/5} 
\frac{\beta^2}{\alpha^{8/5}}
k^{-9} \sum_{i=2,3,4}
(A(C_i)+B(C_i)) ,
\label{Tzetase}
\end{eqnarray}
where we have used eq.(\ref{calP}) in the third equality.
The corresponding nonlinear parameter $\tau_{NL}(k,C_2,C_3,C_4)$ is 
\begin{eqnarray}
\tau_{NL}^{se}(k,C_2,C_3,C_4)&\!\!\equiv\!\!& 
\frac{k^9}{\sqrt{2}}
\left[ \sum_{i=2,3,4}(1+C_i)^{-3/2} \right]^{-1} 
\frac{T_{\zeta,se}(k,C_2,C_3,C_4)}
{(2 \pi^2{\cal P}_{\zeta}(k_0))^{3}}
 \nonumber\\
&\eq&
\frac{k^9}{\sqrt{2}}
(2 \pi^2{\cal P}_{\zeta}(k))^{-3}
\frac{2^7\sqrt{2}}{\pi}\left(\Gamma( 1/4) \right)^{-4}\beta^2 \alpha^{-13/4} 
\left(\frac{H}{M}\right)^{11/2} k^{-9}
\nonumber\\
&&\qquad\qquad\qquad\qquad\qquad\qquad\qquad\times
\sum_{i=2,3,4}(A(C_i)+B(C_i)) 
\left[ \sum_{i=2,3,4}(1+C_i)^{-3/2} \right]^{-1} 
 \nonumber\\
&\eq&
\frac{2^4}{ \pi^7(\Gamma( 1/4 ))^{4}}\frac{\beta^2}{ \alpha^{13/4}}
\left(\frac{H}{M}\right)^{11/2} 
({\cal P}_{\zeta}(k))^{-3}
\sum_{i=2,3,4}(A(C_i)+B(C_i))\left[ \sum_{i=2,3,4}(1+C_i)^{-3/2} \right]^{-1} .
\end{eqnarray}

Substituting eq.~(\ref{calP}) to this expression, we obtain
\begin{eqnarray}
\tau_{NL}^{se}(k,C_2,C_3,C_4)&\!\!\simeq \!\!&
\frac{2^4}{ \pi^7\left(\Gamma( 1/4) \right)^{4}}\frac{\beta^2}{ \alpha^{13/4}}
\left(\pi (\left(\Gamma( 1/4) \right)^{2})\alpha^{3/4}{\cal P}_{\zeta}(k) \right)^{11/5}
({\cal P}_{\zeta}(k))^{-3}
\nonumber\\
&&\qquad\qquad\qquad\qquad\qquad\qquad\qquad\times
\left[\sum_{i=2,3,4}(A(C_i)+B(C_i))\right]\left[ \sum_{i=2,3,4}(1+C_i)^{-3/2} \right]^{-1}
\nonumber\\
&\eq&
\frac{\beta^2}{\alpha^{8/5}} ({\cal P}_{\zeta}(k))^{-4/5}
\left( 2^4\pi^{-24/5}\left(\Gamma( 1/4) \right)^{2/5}
\right) \left[\sum_{i=2,3,4}(A(C_i)+B(C_i))\right]
\left[ \sum_{i=2,3,4}(1+C_i)^{-3/2} \right]^{-1}
 \nonumber\\
&\!\!\simeq \!\!&
8.945 \times10^5\times
\frac{\beta^2}{\alpha^{8/5}} 
\left(\frac{({\cal P}_{\zeta}(k))^{1/2}}{4.8\times10^{-5}}\right)^{-8/5}
\left[\sum_{i=2,3,4}(A(C_i)+B(C_i))\right]\left[ \sum_{i=2,3,4}(1+C_i)^{-3/2} \right]^{-1} .
\label{tause}
\end{eqnarray}


\begin{thebibliography}{99}

\bibitem{Komatsu:2010fb}
  E.~Komatsu {\it et al.},
  arXiv:1001.4538 [astro-ph.CO].

\bibitem{GW}
  U.~Seljak and M.~Zaldarriaga,
  Phys.\ Rev.\ Lett.\  {\bf 78}, 2054 (1997)
  [arXiv:astro-ph/9609169].

  M.~Zaldarriaga and U.~Seljak,
  Phys.\ Rev.\  D {\bf 55}, 1830 (1997)
  [arXiv:astro-ph/9609170].

  N.~Seto, S.~Kawamura and T.~Nakamura,
  Phys.\ Rev.\ Lett.\  {\bf 87}, 221103 (2001)
  [arXiv:astro-ph/0108011].

  G.~M.~Harry, P.~Fritschel, D.~A.~Shaddock, W.~Folkner and E.~S.~Phinney,
  Class.\ Quant.\ Grav.\  {\bf 23}, 4887 (2006)
  [Erratum-ibid.\  {\bf 23}, 7361 (2006)].

\bibitem{Maldacena:2002vr}
  J.~M.~Maldacena,
  JHEP {\bf 0305}, 013 (2003)
  [arXiv:astro-ph/0210603].

\bibitem{squeeze}
  A.~D.~Linde and V.~F.~Mukhanov,
  Phys.\ Rev.\  D {\bf 56}, 535 (1997)
  [arXiv:astro-ph/9610219].

  N.~Bartolo, S.~Matarrese and A.~Riotto,
  Phys.\ Rev.\  D {\bf 65}, 103505 (2002)
  [arXiv:hep-ph/0112261].

  F.~Bernardeau and J.~P.~Uzan,
  Phys.\ Rev.\  D {\bf 66}, 103506 (2002)
  [arXiv:hep-ph/0207295].

  F.~Bernardeau and J.~P.~Uzan,
  Phys.\ Rev.\  D {\bf 67}, 121301 (2003)
  [arXiv:astro-ph/0209330].

  G.~Dvali, A.~Gruzinov and M.~Zaldarriaga,
  Phys.\ Rev.\  D {\bf 69}, 023505 (2004)
  [arXiv:astro-ph/0303591].

  P.~Creminelli,
  JCAP {\bf 0310}, 003 (2003)
  [arXiv:astro-ph/0306122].

  A.~Gruzinov,
  Phys.\ Rev.\  D {\bf 71}, 027301 (2005)
  [arXiv:astro-ph/0406129].

  N.~Bartolo, E.~Komatsu, S.~Matarrese and A.~Riotto,
  Phys.\ Rept.\  {\bf 402}, 103 (2004)
  [arXiv:astro-ph/0406398].

  K.~Enqvist, A.~Jokinen, A.~Mazumdar, T.~Multamaki and A.~Vaihkonen,
  Phys.\ Rev.\ Lett.\  {\bf 94}, 161301 (2005)
  [arXiv:astro-ph/0411394].

  D.~Seery and J.~E.~Lidsey,
  JCAP {\bf 0506}, 003 (2005)
  [arXiv:astro-ph/0503692].

  D.~Seery and J.~E.~Lidsey,
  JCAP {\bf 0509}, 011 (2005)
  [arXiv:astro-ph/0506056].

  A.~Jokinen and A.~Mazumdar,
  JCAP {\bf 0604}, 003 (2006)
  [arXiv:astro-ph/0512368].

  D.~H.~Lyth,
  JCAP {\bf 0511}, 006 (2005)
  [arXiv:astro-ph/0510443].

  M.~P.~Salem,
  Phys.\ Rev.\  D {\bf 72}, 123516 (2005)
  [arXiv:astro-ph/0511146].

  M.~Sasaki, J.~Valiviita and D.~Wands,
  Phys.\ Rev.\  D {\bf 74}, 103003 (2006)
  [arXiv:astro-ph/0607627].

  K.~A.~Malik and D.~H.~Lyth,
  JCAP {\bf 0609}, 008 (2006)
  [arXiv:astro-ph/0604387].

  N.~Barnaby and J.~M.~Cline,
  Phys.\ Rev.\  D {\bf 73}, 106012 (2006)
  [arXiv:astro-ph/0601481].

  L.~Alabidi and D.~Lyth,
  JCAP {\bf 0608}, 006 (2006)
  [arXiv:astro-ph/0604569].

  X.~Chen, M.~x.~Huang, S.~Kachru and G.~Shiu,
  JCAP {\bf 0701}, 002 (2007)
  [arXiv:hep-th/0605045].

  X.~Chen, M.~x.~Huang and G.~Shiu,
  Phys.\ Rev.\  D {\bf 74}, 121301 (2006)
  [arXiv:hep-th/0610235].

  X.~Chen, R.~Easther and E.~A.~Lim,
  JCAP {\bf 0706}, 023 (2007)
  [arXiv:astro-ph/0611645].

  L.~Alabidi,
  JCAP {\bf 0610}, 015 (2006)
  [arXiv:astro-ph/0604611].

  D.~Seery, J.~E.~Lidsey and M.~S.~Sloth,
  JCAP {\bf 0701}, 027 (2007)
  [arXiv:astro-ph/0610210].

  C.~T.~Byrnes, M.~Sasaki and D.~Wands,
  Phys.\ Rev.\  D {\bf 74}, 123519 (2006)
  [arXiv:astro-ph/0611075].

  T.~Suyama and M.~Yamaguchi,
  Phys.\ Rev.\  D {\bf 77}, 023505 (2008)
  [arXiv:0709.2545 [astro-ph]].

  F.~Arroja and K.~Koyama,
  Phys.\ Rev.\  D {\bf 77}, 083517 (2008)
  [arXiv:0802.1167 [hep-th]].

  F.~Arroja, S.~Mizuno and K.~Koyama,
  JCAP {\bf 0808}, 015 (2008)
  [arXiv:0806.0619 [astro-ph]].

  M.~Sasaki,
  Prog.\ Theor.\ Phys.\  {\bf 120}, 159 (2008)
  [arXiv:0805.0974 [astro-ph]].

  C.~T.~Byrnes, K.~Y.~Choi and L.~M.~H.~Hall,
  JCAP {\bf 0810}, 008 (2008)
  [arXiv:0807.1101 [astro-ph]].

  C.~T.~Byrnes, K.~Y.~Choi and L.~M.~H.~Hall,
  JCAP {\bf 0902}, 017 (2009)
  [arXiv:0812.0807 [astro-ph]].

  B.~Dutta, L.~Leblond and J.~Kumar,
  Phys.\ Rev.\  D {\bf 78}, 083522 (2008)
  [arXiv:0805.1229 [hep-th]].

  A.~Naruko and M.~Sasaki,
  Prog.\ Theor.\ Phys.\  {\bf 121}, 193 (2009)
  [arXiv:0807.0180 [astro-ph]].

  T.~Suyama and F.~Takahashi,
  JCAP {\bf 0809}, 007 (2008)
  [arXiv:0804.0425 [astro-ph]].

  X.~Gao,
  JCAP {\bf 0806}, 029 (2008)
  [arXiv:0804.1055 [astro-ph]].

  H.~R.~S.~Cogollo, Y.~Rodriguez and C.~A.~Valenzuela-Toledo,
  JCAP {\bf 0808}, 029 (2008)
  [arXiv:0806.1546 [astro-ph]].

  K.~Ichikawa, T.~Suyama, T.~Takahashi and M.~Yamaguchi,
  Phys.\ Rev.\  D {\bf 78}, 023513 (2008)
  [arXiv:0802.4138 [astro-ph]].

  C.~T.~Byrnes,
  JCAP {\bf 0901}, 011 (2009)
  [arXiv:0810.3913 [astro-ph]].

  D.~Langlois, F.~Vernizzi and D.~Wands,
  JCAP {\bf 0812}, 004 (2008)
  [arXiv:0809.4646 [astro-ph]].

  C.~Hikage, K.~Koyama, T.~Matsubara, T.~Takahashi and M.~Yamaguchi,
  Mon.\ Not.\ Roy.\ Astron.\ Soc.\  {\bf 398}, 2188 (2009)
  [arXiv:0812.3500 [astro-ph]].

  M.~Kawasaki, K.~Nakayama, T.~Sekiguchi, T.~Suyama and F.~Takahashi,
  JCAP {\bf 0811}, 019 (2008)
  [arXiv:0808.0009 [astro-ph]].

  Q.~G.~Huang,
  JCAP {\bf 0811}, 005 (2008)
  [arXiv:0808.1793 [hep-th]].

  X.~Gao,
  arXiv:0904.4187 [hep-th].

  Q.~G.~Huang,
  JCAP {\bf 0905}, 005 (2009)
  [arXiv:0903.1542 [hep-th]].

  Q.~G.~Huang,
  JCAP {\bf 0906}, 035 (2009)
  [arXiv:0904.2649 [hep-th]].

  C.~T.~Byrnes and G.~Tasinato,
  JCAP {\bf 0908}, 016 (2009)
  [arXiv:0906.0767 [astro-ph.CO]].

  C.~Cheung, A.~L.~Fitzpatrick, J.~Kaplan and L.~Senatore,
  JCAP {\bf 0802}, 021 (2008)
  [arXiv:0709.0295 [hep-th]].

\bibitem{Chen:2006nt}
  X.~Chen, M.~x.~Huang, S.~Kachru and G.~Shiu,
  JCAP {\bf 0701}, 002 (2007)
  [arXiv:hep-th/0605045].


\bibitem{equilateral}
  M.~Alishahiha, E.~Silverstein and D.~Tong,
  Phys.\ Rev.\  D {\bf 70}, 123505 (2004)
  [arXiv:hep-th/0404084].


  D.~Langlois, S.~Renaux-Petel, D.~A.~Steer and T.~Tanaka,
  Phys.\ Rev.\ Lett.\  {\bf 101}, 061301 (2008)
  [arXiv:0804.3139 [hep-th]].

  D.~Langlois, S.~Renaux-Petel, D.~A.~Steer and T.~Tanaka,
  Phys.\ Rev.\  D {\bf 78}, 063523 (2008)
  [arXiv:0806.0336 [hep-th]].

  S.~Li, Y.~F.~Cai and Y.~S.~Piao,
  Phys.\ Lett.\  B {\bf 671}, 423 (2009)
  [arXiv:0806.2363 [hep-ph]].

  X.~Gao and B.~Hu,
  JCAP {\bf 0908}, 012 (2009)
  [arXiv:0903.1920 [astro-ph.CO]].

  Y.~F.~Cai and H.~Y.~Xia,
  Phys.\ Lett.\  B {\bf 677}, 226 (2009)
  [arXiv:0904.0062 [hep-th]].

  D.~Langlois, S.~Renaux-Petel and D.~A.~Steer,
  JCAP {\bf 0904}, 021 (2009)
  [arXiv:0902.2941 [hep-th]].

  J.~Khoury and F.~Piazza,
  JCAP {\bf 0907}, 026 (2009)
  [arXiv:0811.3633 [hep-th]].

  S.~Mizuno, F.~Arroja, K.~Koyama and T.~Tanaka,
  Phys.\ Rev.\  D {\bf 80}, 023530 (2009)
  [arXiv:0905.4557 [hep-th]].

  M.~x.~Huang, G.~Shiu and B.~Underwood,
  Phys.\ Rev.\  D {\bf 77}, 023511 (2008)
  [arXiv:0709.3299 [hep-th]].


\bibitem{ArkaniHamed:2003uz}
  N.~Arkani-Hamed, P.~Creminelli, S.~Mukohyama and M.~Zaldarriaga,
  JCAP {\bf 0404}, 001 (2004)
  [arXiv:hep-th/0312100].

\bibitem{Holman:2007na}
  R.~Holman and A.~J.~Tolley,
  JCAP {\bf 0805}, 001 (2008)
  [arXiv:0710.1302 [hep-th]].


\bibitem{trispectrum}
  D.~Seery and J.~E.~Lidsey,
  JCAP {\bf 0701}, 008 (2007)
  [arXiv:astro-ph/0611034].

  D.~Seery, M.~S.~Sloth and F.~Vernizzi,
  JCAP {\bf 0903}, 018 (2009)
  [arXiv:0811.3934 [astro-ph]].

  Y.~Rodriguez and C.~A.~Valenzuela-Toledo,
  Phys.\ Rev.\  D {\bf 81}, 023531 (2010)
  [arXiv:0811.4092 [astro-ph]].

  X.~Chen, B.~Hu, M.~x.~Huang, G.~Shiu and Y.~Wang,
  JCAP {\bf 0908}, 008 (2009)
  [arXiv:0905.3494 [astro-ph.CO]].

  X.~Gao, M.~Li and C.~Lin,
  JCAP {\bf 0911}, 007 (2009)
  [arXiv:0906.1345 [astro-ph.CO]].


\bibitem{Chen:2009bc}
  X.~Chen, B.~Hu, M.~x.~Huang, G.~Shiu and Y.~Wang,
  JCAP {\bf 0908}, 008 (2009)
  [arXiv:0905.3494 [astro-ph.CO]].

\bibitem{Huang:2006eha}
  X.~Chen, M.~x.~Huang and G.~Shiu,
  Phys.\ Rev.\  D {\bf 74}, 121301 (2006)
  [arXiv:hep-th/0610235].

\bibitem{Arroja:2009pd}
  F.~Arroja, S.~Mizuno, K.~Koyama and T.~Tanaka,
  Phys.\ Rev.\  D {\bf 80}, 043527 (2009)
  [arXiv:0905.3641 [hep-th]].
  
\bibitem{Mizuno:2009mv}
  S.~Mizuno, F.~Arroja and K.~Koyama,
  Phys.\ Rev.\  D {\bf 80}, 083517 (2009)
  [arXiv:0907.2439 [hep-th]].

\bibitem{RenauxPetel:2009sj}
  S.~Renaux-Petel,
   ``Combined local and equilateral non-Gaussianities from multifield DBI
  JCAP {\bf 0910}, 012 (2009)
  [arXiv:0907.2476 [hep-th]].











\bibitem{ArkaniHamed:2003uy}
  N.~Arkani-Hamed, H.~C.~Cheng, M.~A.~Luty and S.~Mukohyama,
  JHEP {\bf 0405}, 074 (2004)
  [arXiv:hep-th/0312099].

\bibitem{ArkaniHamed:2005gu}
  N.~Arkani-Hamed, H.~C.~Cheng, M.~A.~Luty, S.~Mukohyama and T.~Wiseman,
  JHEP {\bf 0701}, 036 (2007)
  [arXiv:hep-ph/0507120].

\bibitem{Furukawa:2010gr}
  T.~Furukawa, S.~Yokoyama, K.~Ichiki, N.~Sugiyama and S.~Mukohyama,
  arXiv:1001.4634 [astro-ph.CO].


\bibitem{Bardeen:1983qw}
  J.~M.~Bardeen, P.~J.~Steinhardt and M.~S.~Turner,
  Phys.\ Rev.\  D {\bf 28}, 679 (1983).
  
\bibitem{Senatore:2004rj}
  L.~Senatore,
  Phys.\ Rev.\  D {\bf 71}, 043512 (2005)
  [arXiv:astro-ph/0406187].


\bibitem{Creminelli:2005hu}
  P.~Creminelli, A.~Nicolis, L.~Senatore, M.~Tegmark and M.~Zaldarriaga,
  JCAP {\bf 0605}, 004 (2006)
  [arXiv:astro-ph/0509029].


  
\bibitem{Kogo:2006kh}
  N.~Kogo and E.~Komatsu,
  Phys.\ Rev.\  D {\bf 73}, 083007 (2006)
  [arXiv:astro-ph/0602099].
  
\bibitem{Weinberg:2005vy}
  S.~Weinberg,
  Phys.\ Rev.\  D {\bf 72}, 043514 (2005)
  [arXiv:hep-th/0506236].

  


\bibitem{Byrnes:2006vq}
  C.~T.~Byrnes, M.~Sasaki and D.~Wands,
  Phys.\ Rev.\  D {\bf 74}, 123519 (2006)
  [arXiv:astro-ph/0611075].



\bibitem{Huang:2010ab}
  Q.~G.~Huang,
  arXiv:1004.0808 [astro-ph.CO].






\end{thebibliography}
\end{document}